\documentclass[aps,prl,reprint,groupedaddress,a4paper,amsmath,preprintnumbers]{revtex4-2}

\usepackage[colorlinks]{hyperref}
\usepackage[english]{babel}
\usepackage[utf8]{inputenc}
\usepackage{orcidlink}
\usepackage{mathtools}
\usepackage{physics}
\usepackage{chemformula}
\usepackage[group-minimum-digits=0,table-figures-decimal=0,table-number-alignment=center]{siunitx}
\usepackage{qcircuit}
\usepackage{import}

\usepackage{tikz}
\usetikzlibrary{calc}

\newcommand{\latin}[1]{{\emph{#1}}}

\def\equationautorefname~#1\null{Eq.\,(#1)\null}

\usepackage{dsfont}

\newcommand{\Z}{\mathds{Z}}

\newcommand{\bigO}[1]{\ensuremath{\mathop{}\mathopen{}O\mathopen{}\left(#1\right)}} 
 
\newcommand{\ceil}[1]{{\ensuremath{\left\lceil#1\right\rceil}}}
\newcommand{\floor}[1]{{\ensuremath{\left\lfloor#1\right\rfloor}}}
\newcommand{\identite}{\ensuremath{\mathds{1}}}

\makeatletter
\def\input@path{{../graphiques/}{../graphiques/arithmetique/}}
\makeatother

\usepackage[caption=false]{subfig}
\begin{document}

\title{Factoring \num[detect-all]{2048}-bit RSA Integers in \num[detect-all]{177}~Days with \num[detect-all]{13436}~Qubits \\ and a Multimode Memory}

\author{Élie Gouzien\,\orcidlink{0000-0002-8209-0681}}
\email[]{elie.gouzien@cea.fr}
\author{Nicolas Sangouard\,\orcidlink{0000-0002-3136-0266}}
\homepage[]{https://quantum.paris}
\affiliation{Université Paris--Saclay, CEA, CNRS, Institut de Physique Théorique, \num{91191} Gif-sur-Yvette, France}

\date{\today}

\preprint{\href{https://arxiv.org/abs/2103.06159}{arXiv:2103.06159}}
\doi{10.1103/PhysRevLett.127.140503}

\begin{abstract}
We analyze the performance of a quantum computer architecture combining a small processor and a storage unit.
By focusing on integer factorization, we show a reduction by several orders of magnitude of the number of processing qubits compared with a standard architecture using a planar grid of qubits with nearest-neighbor connectivity.
This is achieved by taking advantage of a temporally and spatially multiplexed memory to store the qubit states between processing steps.
Concretely, for a characteristic physical gate error rate of $10^{-3}$, a processor cycle time of 1~microsecond, factoring a \num{2048}-bit RSA integer is shown to be possible in \num{177}~days with 3D gauge color codes assuming a threshold of \SI{0.75}{\percent} with a processor made with \num{13436}~physical qubits and a memory that can store \num{28}~million spatial modes and \num{45}~temporal modes with 2~hours' storage time.
By inserting additional error-correction steps, storage times of 1~second are shown to be sufficient at the cost of increasing the run-time by about \SI{23}{\percent}.
Shorter run-times (and storage times) are achievable by increasing the number of qubits in the processing unit.
We suggest realizing such an architecture using a microwave interface between a processor made with superconducting qubits and a multiplexed memory using the principle of photon echo in solids doped with rare-earth ions.
\end{abstract}

\maketitle

\addtolength{\parskip}{\bigskipamount}

\paragraph{Introduction ---}
Superconducting qubits form building blocks of one of the most advanced platforms for realizing quantum computers~\cite{OliverARoCMP2020SuperconductingQubitsCurrent,SolanoAiPX2018Digitalanalogquantum}.
The standard architecture consists of laying superconducting qubits in a 2D grid and computing using only neighboring interactions.
Recent estimations showed however that fault-tolerant realizations of various quantum algorithms with this architecture would require millions of physical qubits~\cite{Ekeraa2019Howfactor2048, Babbush2020CompilationFaultTolerant, Babbush2020Evenmoreefficient}.
These performance analyses naturally raise the question of an architecture better exploiting the potential of superconducting qubits.

In developing a quantum architecture we have much to learn from classical architectures.
Realizations using trapped ions for example combine processing with storage units~\cite{WinelandN2002Architecturelargescale}.
The authors of Ref.\,\cite{Chong2006QuantumMemoryHierarchies} realized that key quantum algorithms are mostly sequential meaning that we may only need a small computing block for all the qubits in the storage unit in this architecture.
Ongoing experimental efforts aim at exploiting this idea to reduce the number of superconducting qubits in the standard approach to quantum computing by adding a quantum memory implemented with spins or atoms~\cite{NoriRoMP2013Hybridquantumcircuits,SchmiedmayerPotNAoS2015Quantumtechnologieshybrid,BertetCRP2016Towardsspinensemble}.
A detailed analysis of the performance of this hybrid architecture is however missing.

We here report on such an analysis by considering a quantum memory that can store multiple spatial transverse and temporal modes.
The memory can be thought of as a qubit register in which the address of each qubit is identified by a temporal and a spatial index.
When a given qubit needs to be processed, its state is released and mapped into the processor by means of a microwave field in a temporal and spatial mode corresponding to the qubit address.
When the processing is done, the qubit state is mapped back to the memory and stored until another processing operation is needed.

\begin{figure}
\includegraphics[width=\linewidth]{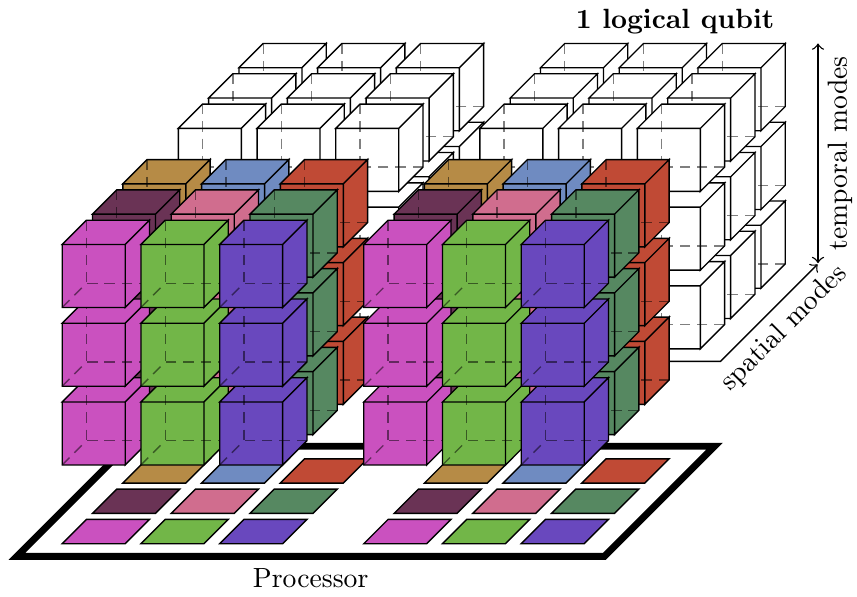}
\caption{Quantum computer architecture using a processor made with a 2D grid of qubits and a memory operating as a qubit register where the address of each qubit is specified by a temporal and spatial index.
Only (dressed) logical qubits are represented; additional ancillary qubits are used for measuring the operators for error correction.
}\label{fig:architecture}
\end{figure}

More precisely, we use 3D error-correction codes~\cite{Martin-DelgadoPRB2007Exacttopologicalquantum} in which the address of each (dressed) logical qubit is encoded into a 3D structure of physical addresses, two dimensions being encoded in space and one in time (see~\autoref{fig:architecture}).
Error-correction and logical gates are applied by sequentially releasing physical qubits corresponding to different ``horizontal'' slices (with different temporal indexes) and by processing each slice (with the same temporal indexes) simultaneously.

We assess the performance of this architecture through a version of Shor's algorithm~\cite{Shor1994Algorithmsquantumcomputation} proposed by Ekerå and Håstad~\cite{Haastad2017QuantumAlgorithmsComputing}.
The algorithm is a threat for widely used cryptosystems based either on the factorization~\cite{AdlemanCotA1978methodobtainingdigital} or the discrete logarithm problem~\cite{HellmanIToIT1976Newdirectionscryptography,nullITL2013DigitalSignatureStandard}.
It can also be considered as a certification tool to check the proper functioning of an actual quantum computer as its outcome can be verified efficiently.
Last but not least, the cost of its implementation has been evaluated using plausible physical assumptions for a large scale processor with a standard 2D grid of superconducting qubits (a characteristic physical gate error rate of $10^{-3}$, a surface code cycle time of \SI{1}{\micro\second}, and a reaction time of \SI{10}{\micro\second}): it was estimated that it should be possible to factor a \num{2048}-bit integer, typically used in the Rivest--Shamir--Adleman (RSA) cryptosystem, in \num{8}~hours with \num{20}~million qubits~\cite{Ekeraa2019Howfactor2048}.

By taking this estimation as a reference, we estimate the cost of implementing the same version of Shor's algorithm 
in terms of physical processing qubit number, multimode capacity, memory storage time, and run-time.
Our evaluation is given in the case where the processor is made with two (dressed) logical qubit slices.
Under the assumptions used in Ref.\,\cite{Ekeraa2019Howfactor2048} for the gate error rate and the cycle time, we show that it should be possible to factor a \num{2048}-bit RSA integer in \num{177}~days using a multimode memory with a storage time of about \num{2}~hours and a processor including \num{13436}~physical qubits---a reduction by more than 3~orders of magnitude of the number of physical qubits, as compared to the standard architecture without memory~\cite{Ekeraa2019Howfactor2048}, at the cost of a $\approx 500$ times longer run-time.
By inserting additional error-correction steps, we show that the storage time can be significantly reduced at the cost of a slight increase of run-time.
We also explain how shorter run-times and storage times are achievable at the cost of increasing the number of qubits in the processing unit.
We propose a realization of such an architecture using a microwave interface between a processor made with superconducting qubits and a multiplexed memory using the principle of photon echo in solids doped with rare-earth ions embedded in cavities.

\paragraph{Principles of (a variant of) Shor's algorithm ---}
Consider the factorization of $N = p \times q$, the product of two prime numbers of similar sizes, $p$ and $q$.
We note $n$ the number of bits involved in the binary representation of $N$, that is $2^{n-1} \leq N < 2^{n}$.
While no efficient classical factorization algorithm is known, Shor's algorithm and its variants factor $N$ with a polynomial complexity into $n$~\cite{Shor1994Algorithmsquantumcomputation,ShorSJoC1997PolynomialTimeAlgorithms,Ekeraa2016ModifyingShorsalgorithm,Ekeraa2017postprocessingquantum,Haastad2017QuantumAlgorithmsComputing,Ekeraa2018Quantumalgorithmscomputing}.

The version of Shor's factorization algorithm proposed by Ekerå and Håstad~\cite{Haastad2017QuantumAlgorithmsComputing}
starts by randomly selecting an integer $g$ in the multiplicative group of integers modulo $N$, $\Z_{N}^*$, and defining $h = g^{{(N-1)}/2}$. 
As the order of $\Z_{N}^*$ is $\phi(N) = (p-1) (q-1)$, we have $h = g^{(pq-p-q+1)/2}g^{(p+q-2)/2} \equiv g^{(p+q-2)/2} \mod{N}$ where the last equivalence is the result of the Chinese remainder theorem.
Under the assumption that the order $r$ of $g$ (the smallest non-negative integer such that $g^r \equiv 1 \mod N$) satisfies $r > {(p + q - 2)}/2$, computing the discrete logarithm of $h$ modulo $N$, as detailed later, yields $l = {(p + q - 2)}/2$.
For large $N$, the assumption is verified with a high probability~\cite{Haastad2017QuantumAlgorithmsComputing}.
Using $N = pq$ and $l = {(p + q - 2)}/2$, where $N$ and $l$ are both known, $p$ and $q$ are recovered by choosing one solution of the equation $N = p(2l+2-p)$, and then exploiting $q = 2l + 2 - p$.

The discrete logarithm is computed in three steps.
First, the exponentiation $(e_1,e_2) \rightarrow g^{e_1} h^{-e_2}$ is applied once on two quantum registers prepared in a superposition of every possible value of $e_1$ and $e_2$, respectively.
Two quantum Fourier transforms are then applied independently to the two registers before being measured.
Finally, a classical postprocessing extracts the discrete logarithm $l$ of $h$ modulo $N$ from the measurement results.
Because the measurements are performed directly after the Fourier transform, the cost of exponentiation largely dominates the cost of Ekerå and Håstad's algorithm (see \autoref{appendix:qft}).

\paragraph{Number of gates ---}
The modular exponentiation needed in Ekerå and Håstad's algorithm, \latin{i.e.\@}, the operation $\ket{e}\ket{1} \mapsto \ket{e}\ket{g^e \mod N}$, with the input $e$ and the output $g^e \mod N$ encoded on $n_e$ and $n$~bits, respectively, can be decomposed into $n_e$~multiplications, each being decomposed into $2 n$~controlled additions of integers of typical size $n$ and one controlled swap between two registers of size $n$, giving a total number of $2 n_e n$ ($n_e$) controlled additions (swaps between registers, respectively) (see \autoref{appendix:exponentiation} for details).
Each modular addition is obtained with a standard adder circuit at the cost of a specific representation---the coset representation (see \autoref{appendix:coset_representation})---adding $m$~additional qubits to the register.
A controlled swap operation between two qubits can be performed using two controlled NOTs (CNOTs) and one Toffoli gate.
Hence, the total cost for controlled swaps operating on two registers using $n+m$~qubits is of $2(n+m)$~CNOTs and $n+m$~Toffoli gates (see \autoref{appendix:exponentiation}).
For the controlled addition, we can use a semi-classical adder whose mean cost for integers of size $n+m$ is of $5.5(n+m)-9$~CNOTs and $2(n+m)-1$~Toffoli gates (see \autoref{appendix:exponentiation}).
Given the number of gates in controlled addition and swap operations, the number of additions and swaps in the multiplication, and the number of multiplications in the modular exponentiation, the cost of factorization can easily be estimated (see \autoref{appendix:exponentiation}).
This cost can however be reduced using windowed arithmetic circuits~\cite{Gidney2019Windowedquantumarithmetic}.
The basic idea consists of grouping the bits of $e$ by blocks (each including $w_e$~bits) for controlling each multiplication, hence reducing the number of these multiplications.
Similarly, for each multiplication input bits are grouped (in blocks including $w_m$~bits) to reduce the number of additions composing it.
As detailed in \autoref{appendix:windowed_arithmetic}, the cost of exponentiation is dominated in this case by $2 \frac{n_e (n+m) n}{w_e w_m}$~1-qubit gates, $\left[2^{w_e + w_m}n + 12(n+m)\right] \frac{n_e (n+m)}{w_e w_m}$~CNOTs, and $4 \frac{n_e {(n+m)}^2}{w_e w_m}$~Toffoli gates.
We emphasize that this is a first order estimation.
In the code used to compute the required resources and find optimal parameters, the complete formulae have been used~\footnote{Code is available at \url{https://github.com/ElieGouzien/factoring_with_memory}}.

\paragraph{Error correction ---}
The error correction is achieved using 3D gauge color codes, a family of subsystem codes~\cite{Martin-DelgadoPRB2007Exacttopologicalquantum}.
A first code admits a transversal implementation of CNOT and Hadamard gates while a second code accepts a transversal implementation of the non-Clifford $T$ gate.
Switching between the two codes gives a universal set of gates without the need for state distillation~\cite{BombinNJoP2015Gaugecolorcodes}, contrary to standard ways of operating the surface code~\cite{VuillotN2017Roadstowardsfault}.

The two codes are based on a shared geometrical structure: a large tetrahedron constructed from elementary tetrahedrons (see \autoref{appendix:err_correction} for details).
A physical qubit is attributed to each elementary tetrahedron.
As in any subsystem codes, the stabilized subspace is split into a tensorial product of the (bare) logical and gauge qubits (the dressed logical qubit includes the bare logical qubit and gauge qubits).
A set of operators---generators of gauge operators---are measured, each being the product of (up to six) $X$ (or $Z$) operators associated to qubits corresponding to tetrahedrons sharing the same edge.
From these measurements, the values of stabilizers of the two codes are deduced.
In the code used for implementing $H$ and CNOT gates, the stabilizers are defined from the vertices, \latin{i.e.\@}, the product of $X$ (or $Z$) operators associated to qubits corresponding to tetrahedrons sharing the same vertex.
In the code used for implementing $T$ gates, the stabilizers are defined from the vertices for $X$ operators and from the edges for $Z$ operators.
The value of an operator represented by a vertex is classically recovered by multiplying the measurement results of combinations of specific edges ending at the given vertex. 
Several combinations are possible giving redundancies that can be exploited to achieve fault-tolerant error correction with only one run of measurements~\cite{BombinPRX2015SingleShotFault}.
The structure of codes in which the stabilized subsystem is the tensor product of the gauge and (bare) logical subsystems guarantees that measurements of gauge operators do not reveal the value of the (bare) logical qubit (see \autoref{appendix:err_correction}).

To account for the additional resource needed to implement these codes, we use an estimation of the residual error probability on one logical qubit given in~\cite[eq.\,(4)]{BrowneNC2016Faulttoleranterror}
\begin{equation}\label{eq:logical_error}
p_{\text{logical}}
	= A \exp[\alpha \log(\frac{p}{p_{\text{th}}}) d^{\beta}]
\end{equation}
where $A \approx 0.033$, $\alpha \approx 0.516$, $\beta \approx 0.822$, $p$ is the error probability per physical qubit, $d$ the code distance which is related to the number of physical qubits per logical qubits (see below) and $p_{\text{th}}$ the fault-tolerance threshold.
While the circuit-level threshold  is unknown, we choose $p_{\text{th}} = \SI{0.75}{\percent}$ as a working hypothesis and give in \autoref{appendix:code:threshold} the run-time and the resource as a function of the code threshold.

\paragraph{Architecture ---}
For simplicity, the tetrahedral structure of the error correction (see \autoref{appendix:err_correction}) can be included into a large cube in which physical qubits are now represented by elementary cubes (see~\autoref{fig:architecture}).
The large cubes are stored into the memory and loaded by slices into the processor when they need to be processed.
We size the processor such that one slice of two large cubes can be loaded simultaneously, which is convenient to perform 2-qubit gates efficiently.
Each gate is immediately followed by an error-correction round on the processed qubits.
This is done by reloading again each slice sequentially in the processor and by measuring the gauge generators (before recovering classically the code stabilizers), each of them using up to six 2-qubit gates, one auxiliary qubit and one measurement of this auxiliary~\cite{BombinNJoP2015Gaugecolorcodes,NemotoRoPiP2013Quantumerrorcorrection}.
Note that the codes of interest are 3D local and the auxiliary qubits only need to keep coherence for the time of loading and measuring two successive slices for successfully performing a stabilizer measurement.
Once the syndromes are obtained and the errors are detected, the correction of these errors is delayed and merged with the next operation applied on the qubit to be corrected.
Further note that all-to-all connectivity between the logical qubits is achieved if each physical address in the memory can be mapped to three physical qubits in the processor: two for the 2-qubit gates (depending on whether the physical qubit is the logical control or target qubits) and one for the error correction.
For achieving a code distance $d$ the number of physical qubits in the processor is $n_{\text{qubits}}=2 \times 2 \times \frac{3d^2 + 2d - 3}{2}$, corresponding to two logical qubit slices (see \autoref{appendix:err_correction}) and including the ancillary qubits (essentially one per physical qubit) needed for stabilizer measurements.
For a code distance $d$, we approximate the time it takes to perform one (1-qubit or 2-qubit) logical gate by $2 (d-2) t_c$ where $t_c$ is the cycle time of the 2D processor (time to load one qubit slice; to measure the stabilizers, which is longer than the gate operation; and to reload the slice into the memory) and the factor $2$ comes from the fact that the gate is immediately followed by an error-correction round.

\paragraph{Cost evaluation ---}
To evaluate the resources required for integer factorization, we consider the total number of gates involved in the logical circuit.
The total run-time for one attempt is obtained by multiplying the gate number by the time it takes to perform one gate, while the success probability is deduced from the logical error probability [\autoref{eq:logical_error}].
Following Ref.\,\cite{Ekeraa2019Howfactor2048}, we consider a cycle time of $t_c=\SI{1}{\micro\second}$ and a mean error per physical qubit and per gate of $p=10^{-3}$.
Note that the mean error per gate now includes errors during reading of and writing into the memory.

The cost evaluation is finally obtained by optimizing the two window parameters $w_m$ and $w_e$, the coset representation padding $m$ and the code distance $d$
in order to minimize the volume $t_{\text{exp}} \times n_{\text{qubits}}$.
$t_{\text{exp}}=\frac{t}{p_s}$ is the average time to obtain the result (several attempts might be necessary), with $t$ the computation time per attempt and $p_s$ the success probability.

\paragraph{Results ---}
The required resources to factor a $n$-bit RSA integer are presented in \autoref{fig:results} and discussed in \autoref{appendix:results}.
\begin{figure}
\centering
\includegraphics[width=0.5\textwidth]{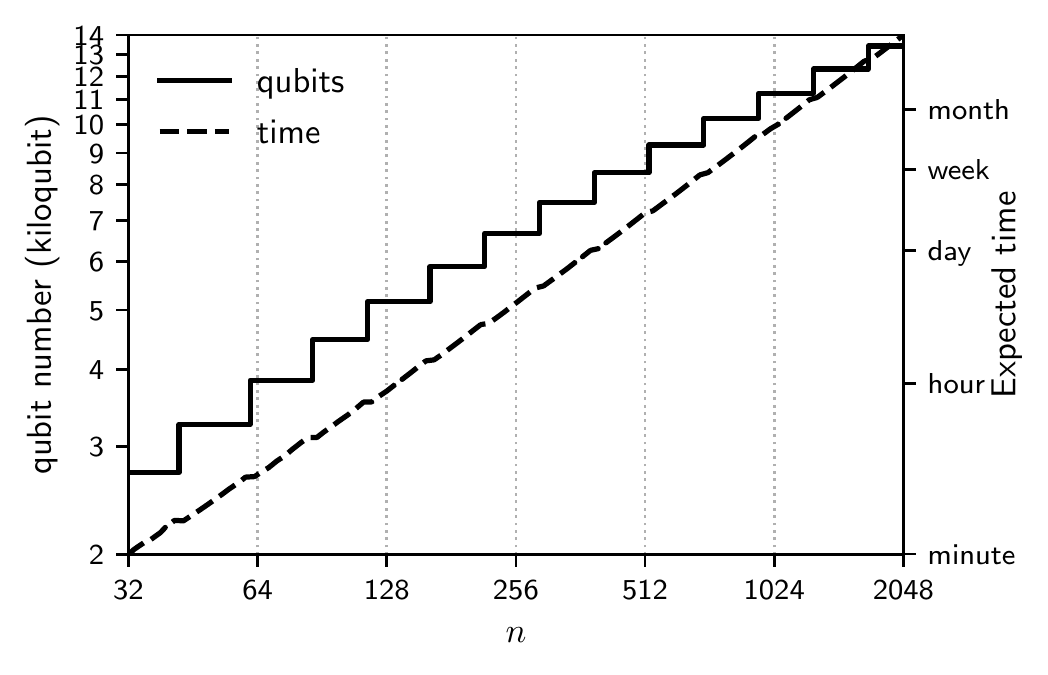}
\caption{Number of qubits in the processor and run-time to factor $n$-bit RSA integers with a computer architecture using a multimode memory.}\label{fig:results}
\end{figure}
Our estimation suggests that the factorization of a \num{2048}-bit integer corresponding to the most common RSA key size would be possible in about \num{177}~days with a processor having only \num{13436}~qubits.
Concerning the memory, we made the hypothesis of an error per cycle of $p=10^{-3}$, including the reading and writing error.
As previously discussed, we need a memory for which each mode can be mapped to three different qubits of the processor.
We estimated the maximum time between storage and readout of the same qubit to be less than 2~hours.
A memory with a storage time of at least 2~hours is however not necessary as error-correction steps can be implemented periodically at the cost of increasing the run-time.
Error correction of all the qubits stored in the memory is estimated to take \SI{186}{\milli\second} with a processor having \num{13436}~bits, meaning that the storage time simply needs to be longer than \SI{186}{\milli\second}.
Applying a correction every second for example would increase the run-time by about \SI{23}{\percent}.
Note also that both the run-time and storage time can be reduced by increasing the size of the processor (see \autoref{appendix:results}).
We also estimated that \num{28}~million spatial modes and \num{45}~temporal modes need to be stored.
Note that the number of stored modes does not enter in the volume and is thus not optimized (see \autoref{appendix:memoire}).
Note also that qubit addresses in the memory can be identified by temporal indexes only at the cost of longer run-time when photon-echo type protocols are used, \latin{cf.\@}, below for a concrete example.

\paragraph{Implementation ---}
Our proposal provides a viable solution to get rid of the individual control of millions of qubits but the challenge now relies on the realization of an efficient multimode quantum memory.
As shown in Ref.\,\cite{WilsonNJoP2013Proposalcoherentquantum}, such a memory could be implemented using a solid-state spin ensemble ($\bar{N}$ spins with an inhomogeneous spectral broadening $\Gamma$), resonantly coupled (with single spin coupling rate $g$) to a frequency tunable single-mode microwave resonator (of length $L$ and with damping rate $\kappa$ to an external transmission line).
The resonator serves to enhance microwave absorption and re-emission by the spins.
In particular, unit efficiency absorption of a microwave field can be realized if the finesse $\mathcal{F}$ of the resonator matches the single-path absorption $\alpha L$ of spins $\mathcal{F} = {(\alpha L)}^{-1}$, \latin{i.e.\@}, if the cooperativity $C = \frac{g^2 \bar{N}}{\kappa \Gamma} = \alpha L \times \mathcal{F} = 1$~\cite{SimonPRA2010Impedancematchedcavity}.
Once absorbed, the microwave field can be re-emitted by time reversing the inhomogeneous dephasing using a spin echo technique~\cite{Sangouard2018QuantumOpticalMemory}.
Detuning the resonator off and on resonance at the right time, the spin coherence is recovered, leading to a noise-free, unit re-emission probability of the stored photon if $C=1$~\cite{WilsonNJoP2013Proposalcoherentquantum}.
In the regime $\kappa \gg g\sqrt{\bar{N}} \gg \Gamma$, the memory bandwidth is given by $4\Gamma$~\cite{WilsonNJoP2013Proposalcoherentquantum}, meaning that any input with a spectrum, say, ten times
thinner \latin{i.e.\@}, $4 \Gamma / 10$ can be stored with close to unit efficiency.
Furthermore, the time duration during which an optical coherence can be preserved is limited by the inverse of the homogeneous linewidth $\gamma_h$~\cite{WilsonNJoP2013Proposalcoherentquantum}.
Assuming that the storage efficiency is unchanged if the storage time is hundred times shorter than $\gamma_h^{-1}$, this means that the number of temporal modes that can be stored with almost unit efficiencies is roughly given by $\Gamma/(250 \gamma_h)$.
Interestingly, a well-identified temporal mode can be released while keeping all the other modes in the memory by appropriately detuning the resonator off and on resonance with the spins at the cost of introducing a dead time between two readouts of half the duration of the stored train of pulses on average.

To give an idea of what could be realized in a near future, we estimate that it should be possible to factor $35$ in about \SI{1}{\min} using the exact algorithm presented here (with windowed arithmetic and 3D color codes) and a setup combining a memory for storing \num{38}~logical qubits (\num{3002}~spatial modes and \num{5}~temporal modes) and a processor with \num{316}~physical qubits (we estimate that more than \num{60000}~qubits would be needed with a standard 2D grid and surface code).   
If instead of using a spatially and temporally multiplexed memory, the qubits are stored in the same spatial mode and are identified by (\num{6650}) temporal addresses only, we evaluate the same factorization to be possible in about \num{1}~day using a memory bandwidth $4 \Gamma=\SI[parse-numbers=false]{2 \pi \times 48}{\mega\hertz}$ and taking into account the corresponding dead time between two memory readouts.
In this case, error correction of all the qubits stored in the memory is estimated to take \SI{132}{\milli\second} meaning the storage time needs to be longer than \SI{132}{\milli\second}.
For a memory bandwidth $4 \Gamma = \SI[parse-numbers=false]{2 \pi \times 120}{\mega\hertz}$, the same factorization would take \num{9}~hours, and error correction is estimated to take \SI{53}{\milli\second}.
As discussed in \autoref{appendix:realization}, these requirements can realistically be met with a realization of the memory protocol described before combining a solid doped with rare-earth and a superconducting microwave resonator~\cite{UstinovPRB2011Ultralowpowerspectroscopy, WilsonJoPBAMaOP2012Couplingerbiumspin, BushevPRL2013AnisotropicRareEarth}.

\paragraph{Conclusion ---}
We have shown that the use of a quantum memory for quantum computing is appealing as unprocessed qubits can be loaded into the memory which significantly reduces the size of the processor compared with standard architectures where all qubits are kept in the processor.
All-to-all connectivity between logical qubits is reached if each address in the memory can be mapped to only 3 qubits in the processor.
The use of a memory allows one to exploit a 3D code on a 2D processor.
If we allow each memory mode to be mapped to any qubit in the processor, all-to-all connectivity between physical qubits can be obtained, hence offering many opportunities for error correction and for implementing algorithms with gates operating between non-neighboring qubits.

\begin{acknowledgments}
We acknowledge M.~Afzelius, J.-D.~Bancal, P.~Bertet, E.~Flurin, P.~Sekatski, X.~Valcarce and J.~Zivy
for stimulating discussions and/or for critically reviewing the manuscript.
We acknowledge funding by the Institut de Physique Théorique (IPhT), Commissariat à l'Énergie Atomique et aux Energies Alternatives (CEA) and the Region Île-de-France in the framework of DIM SIRTEQ\@.
\end{acknowledgments}

\appendix
\setcounter{secnumdepth}{2}

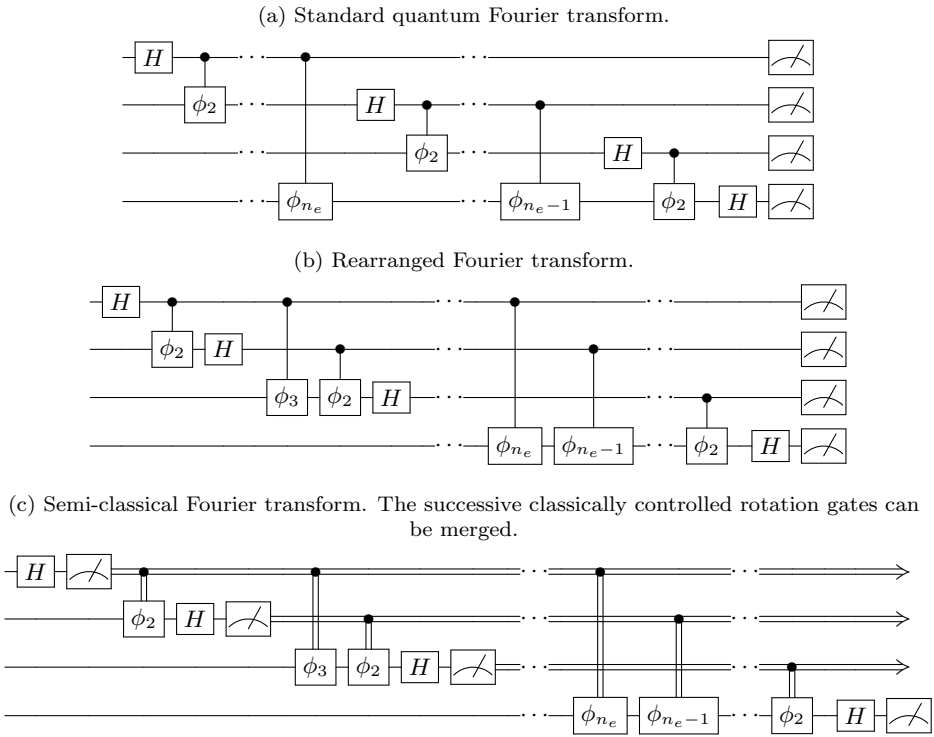
\begin{figure*}
\subfloat[Standard quantum Fourier transform.]{\label{subfig:fourier_transform:quantique}\mbox{
		\small
		\Qcircuit @C=0.5em @R=.5em {
			& \gate{H} & \ctrl{1}      & \push{\cdots} \qw & \ctrl{3}          & \qw & \qw      & \qw           & \push{\cdots} \qw & \qw                 & \qw & \qw      & \qw           & \qw & \qw      & \meter\\
			& \qw      & \gate{\phi_2} & \push{\cdots} \qw & \qw               & \qw & \gate{H} & \ctrl{1}      & \push{\cdots} \qw & \ctrl{2}            & \qw & \qw      & \qw           & \qw & \qw      & \meter\\
			& \qw      & \qw           & \push{\cdots} \qw & \qw               & \qw & \qw      & \gate{\phi_2} & \push{\cdots} \qw & \qw                 & \qw & \gate{H} & \ctrl{1}      & \qw & \qw      & \meter\\
			& \qw      & \qw           & \push{\cdots} \qw & \gate{\phi_{n_e}} & \qw & \qw      & \qw           & \push{\cdots} \qw & \gate{\phi_{n_e-1}} & \qw & \qw      & \gate{\phi_2} & \qw & \gate{H} & \meter
		}}}

\subfloat[Rearranged Fourier transform.]{\label{subfig:fourier_transform:quantique2}\mbox{
		\small
		\Qcircuit @C=0.5em @R=.5em {
			& \gate{H} & \ctrl{1}      & \qw      & \qw & \ctrl{2}      & \qw           & \qw      & \qw & \push{\cdots} \qw & \qw & \ctrl{3}          & \qw                 & \push{\cdots} \qw & \qw           & \qw & \qw      & \meter\\
			& \qw      & \gate{\phi_2} & \gate{H} & \qw & \qw           & \ctrl{1}      & \qw      & \qw & \push{\cdots} \qw & \qw & \qw               & \ctrl{2}            & \push{\cdots} \qw & \qw           & \qw & \qw      & \meter\\
			& \qw      & \qw           & \qw      & \qw & \gate{\phi_3} & \gate{\phi_2} & \gate{H} & \qw & \push{\cdots} \qw & \qw & \qw               & \qw                 & \push{\cdots} \qw & \ctrl{1}      & \qw & \qw      & \meter\\
			& \qw      & \qw           & \qw      & \qw & \qw           & \qw           & \qw      & \qw & \push{\cdots} \qw & \qw & \gate{\phi_{n_e}} & \gate{\phi_{n_e-1}} & \push{\cdots} \qw & \gate{\phi_2} & \qw & \gate{H} & \meter
		}}}

\subfloat[Semi-classical Fourier transform. The successive classically controlled rotation gates can be merged.]{\label{subfig:fourier_transform:semi_classique}\mbox{
		\small
		\Qcircuit @C=0.5em @R=.5em {
			& \gate{H} & \meter & \cctrl{1}     & \cw      & \cw    & \cw & \cctrl{2}      & \cw           & \cw      & \cw    & \cw & \push{\cdots} \cw & \cw & \cctrl{3}         & \cw                 & \push{\cdots} \cw & \cw           & \cw & \cw      & \cwa\\
			& \qw      & \qw    & \gate{\phi_2} & \gate{H} & \meter & \cw & \cw            & \cctrl{1}     & \cw      & \cw    & \cw & \push{\cdots} \cw & \cw & \cw               & \cctrl{2}           & \push{\cdots} \cw & \cw           & \cw & \cw      & \cwa\\
			& \qw      & \qw    & \qw           & \qw      & \qw    & \qw & \gate{\phi_3}  & \gate{\phi_2} & \gate{H} & \meter & \cw & \push{\cdots} \cw & \cw & \cw               & \cw                 & \push{\cdots} \cw & \cctrl{1}     & \cw & \cw      & \cwa\\
			& \qw      & \qw    & \qw           & \qw      & \qw    & \qw & \qw            & \qw           & \qw      & \qw    & \qw & \push{\cdots} \qw & \qw & \gate{\phi_{n_e}} & \gate{\phi_{n_e-1}} & \push{\cdots} \qw & \gate{\phi_2} & \qw & \gate{H} & \meter
		}}}
\caption{
	Different versions of the Fourier transform followed by measurements.
	They are used to convince the reader that the number of gates in the Fourier transform is negligible with respect to the cost of the exponentiation.
	These three versions are based on the phase gates $\phi_k$ defined as a $2\times2$ matrix with diagonal elements $\left(1, e^{\frac{2\pi i}{2^k}}\right)$ and zeros off diagonal. 
	Note that the control and target qubits can be reversed in the representation of each controlled phase gate without changing the result.
	}\label{fig:fourier_transform}
\end{figure*}

\section{Semi-classical Fourier transform}\label{appendix:qft}
We here discuss the semi-classical Fourier transform presented in~\cite{NiuPRL1996SemiclassicalFourierTransform} and show that its cost is negligible.
The standard way to perform the Fourier transform on $n_e$~qubits is shown in
\autoref{subfig:fourier_transform:quantique}: it requires a sequence of one Hadamard and controlled phase gates for each qubit.
In Shor's algorithm as well as in Ekerå and Håstad's version of Shor's algorithm, the qubits are measured right after the Fourier transform, hence explaining the measurements of each qubit at the end of gate sequences in \autoref{subfig:fourier_transform:quantique}.
The simple rearrangement presented in \autoref{subfig:fourier_transform:quantique2} shows that the measurement can be performed right after the Hadamard provided that the following phase gates are classically controlled by the result of this measurement, see \autoref{subfig:fourier_transform:semi_classique}.
In this case, the successive classically controlled phase gates operating on the same qubit can be merged together, leading to a circuit with one phase gate, one Hadamard gate and one measurement per qubit.
When this semi-classical Fourier transform operates on a register made with $n_e$~qubits (the number of bits of the exponent), its cost is linear in $n_e$ and is thus negligible compared to the cubic complexity of the exponentiation.

\section{Decomposition of the exponentiation into elementary gates}\label{appendix:exponentiation}
In this appendix, we aim to give a clear view of how to decompose the modular exponentiation into elementary gates.
The presented method is intended to be simple to understand, but not optimal.
A more efficient one is presented in \autoref{appendix:windowed_arithmetic}.

\subsection{Decomposition of a modular exponentiation into additions}
The modular exponentiation needed in Ekerå and Håstad's algorithm, \latin{i.e.\@}, the operation $\ket{e}\ket{1} \mapsto \ket{e}\ket{g^e \mod N}$, with the input $e$ and the output $g^e \mod N$ encoded on $n_e$ and $n$~bits respectively, can be implemented from controlled modular additions as we show now.
For simplicity, we omit the modulo in this paragraph.

\begin{figure}[h]
	\mbox{\small \providecommand{\controlgate}[1]{\measure{#1}}
\Qcircuit @R=0.4em @C=0.7em {
	\lstick{\ket{e_i}}&\qw           &\qw   &\qw                         &\ctrl{1}                              &\qw     &\rstick{\ket{e_i}}\qw          \\
	\lstick{\ket{x}}  &\qw {/}       &\qw   &\qw                         &\gate{\times g^{2^i}}                 &\qw     &\rstick{\ket{x g^{2^i e_i}}}\qw\\
	                  &              &      &                            &\push{=}                              &        &                               \\
	\lstick{\ket{e_i}}&\qw           &\qw   &\ctrl{1}                    &\ctrl{1}                              &\ctrl{2}&\rstick{\ket{e_i}}\qw          \\
	\lstick{\ket{x}}  &\qw{/}        &\qw   &\controlgate{\text{Input }x}&\gate{+\bar{x}(-g^{-2^i})}            &\qswap  &\rstick{\ket{x g^{2^i e_i}}}\qw\\
	                  &\push{\ket{0}}&\qw{/}&\gate{+x g^{2^i}}\qwx       &\controlgate{\text{Input }\bar{x}}\qwx&\qswap  &\push{\ket{0}}\qw              \\
}
 }
	\caption{
			Principle of a modular multiplication circuit transforming a quantum register encoding the integer $x$ into a state encoding $x \times g^{2^{i} e_i}$.
			A first product-addition operation transforms auxiliary qubits in $\ket{0}$ into $\ket{x \times g^{2^{i}}}$ if $\ket{e_i} = \ket{1}$.
			Then, a product-addition applies $+\bar{x} (-g^{-2^{i}})$ with $\bar{x} = x \times g^{2^{i}}$ into the register encoding $x$ if  $\ket{e_i} = \ket{1}$.
			A final swapping is applied if $\ket{e_i} = \ket{1}$ to put the quantum register into $\ket{x \times g^{2^{i} e_i}}$ and resets the auxiliary qubits to $\ket{0}$.
			Note that all the operations are performed modulo $N$.}\label{fig:multiplication}
\end{figure}
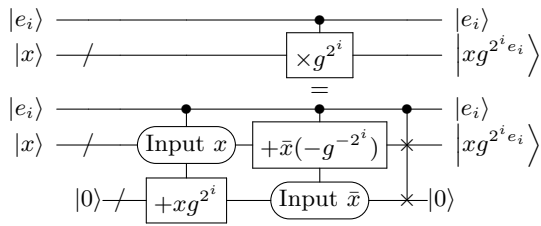

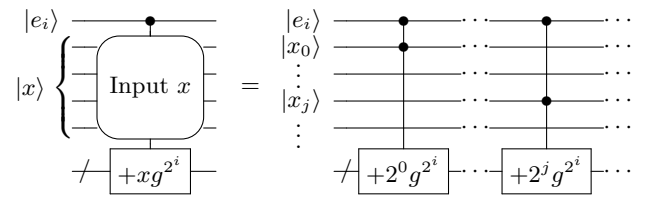
\begin{figure}[h]
	\mbox{\small \providecommand{\multicontrolgate}[2]{\multimeasure{#1}{#2}}
\providecommand{\controlgate}[1]{\measure{#1}}
\newcommand{\multipush}[2]{*+<1em,.9em>{\hphantom{#2}} \POS [0,0]="i",[0,0].[#1,0]="e",!C *{#2},"e"+UR;"e"+UL;"e"+DL;  "e"+DR;"e"+UR,"i"}
\Qcircuit @R=0.2em @C=0.5em {
	\lstick{\ket{e_i}} &\qw   &\ctrl{2}                                &\qw &                &&&&&\lstick{\ket{e_i}}                             &\qw   &\ctrl{2}              &\push{\cdots}\qw&\ctrl{4}              &\push{\cdots}\qw                                           &\\
	                   &      &                                        &    &                &&&&&                                               &      &                      &                &                      &                                                           &\\
	                   &\qw   &\multicontrolgate{3}{\text{Input }x}\qwx&\qw &                &&&&&\lstick{\ket{x_0}}                             &\qw   &\ctrl{5}              &\push{\cdots}\qw&\qw                   &\push{\cdots}\qw                                           &\\
	                   &\qw   &              \ghost{\text{Input }x}    &\qw &\multipush{1}{=}&&&&&\lstick{\raisebox{0.5em}{\vdots\hspace{0.8em}}}&\qw   &\qw                   &\push{\cdots}\qw&\qw                   &\push{\cdots}\qw                                           &\\
	                   &\qw   &              \ghost{\text{Input }x}    &\qw &      \nghost{=}&&&&&\lstick{\ket{x_j}}                             &\qw   &\qw                   &\push{\cdots}\qw&\ctrl{3}              &\push{\cdots}\qw                                           &\\
	                   &\qw   &              \ghost{\text{Input }x}    &\qw &                &&&&&\lstick{\vdots\hspace{0.8em}}                  &\qw   &\qw                   &\push{\cdots}\qw&\qw                   &\push{\cdots}\qw \inputgroupv{3}{6}{0.75em}{1.7em}{\ket{x}}&\\
	                   &      &\qwx                                    &    &                &&&&&                                               &      &                      &                &                      &                                                           &\\
	                   &{/}\qw&\gate{+ x g^{2^i}}\qwx                  &\qw &                &&&&&                                               &{/}\qw&\gate{+ 2^{0} g^{2^i}}&\push{\cdots}\qw&\gate{+ 2^{j} g^{2^i}}&\push{\cdots}\qw                                           &\\
}
 }
	\caption{Decomposition of the first product-addition appearing in each element of the decomposition of the exponentiation into multiplications, see \autoref{fig:multiplication}.}\label{fig:product_add_simple}
\end{figure}

Let $e_{n_e-1} \ldots e_{i} \ldots e_{0}$ be the binary form of $e$.
The exponentiation can first be seen as a sequence of multiplications
\begin{equation}\label{eq:decompose_exponent}
g^e
	= \prod\limits_{i=1}^{n_e-1} g^{2^{i} e_{i}}
	= \prod\limits_{i=1}^{n_e-1} {\left[g^{2^{i}}\right]}^{e_{i}}
\end{equation}
where each multiplication is controlled by the bit value $e_i$.
\autoref{fig:multiplication} shows an implementation of such a multiplication in which a quantum register encoding the integer $x$ ends up into an encoding of $x \times g^{2^{i} e_i}$.
It uses two controlled product-additions, \latin{i.e.\@}, the operation letting $\ket{y}\ket{z}$ unchanged if $\ket{e_i} = \ket{0}$ and mapping $\ket{y}\ket{z}$ into $\ket{y}\ket{z + y \times \gamma}$ ($(y, z, \gamma) \rightarrow (x,0,g^{2^i})$ for the first product-addition appearing in \autoref{fig:multiplication} and $(y, z, \gamma) \rightarrow (\bar{x}, x, - g^{-2^i})$ for the second one, where the negative power stands for multiplicative inverse modulo $N$) when $\ket{e_i} = \ket{1}$.
In case $\ket{e_i} = \ket{1}$, the mapping is performed by considering the binary representation $y_{n-1} \hdots y_0$ of $y$
and by rewriting the product as
\begin{equation}
y \times \gamma
	= \sum\limits_{j=0}^{n-1} \gamma 2^j y_j
	= \sum\limits_{j=0}^{n-1} \left[\gamma 2^j\right] y_j.
\end{equation}
As $y_j$ is either $0$ or $1$, the controlled product-addition can be implemented by a sequence of additions, each of them controlled both by the values of bits $\ket{y_j}$ and $\ket{e_i}$.
\autoref{fig:product_add_simple} shows explicitly the decomposition of the first product-addition appearing into each multiplication of the exponentiation.

We deduce that the modular exponentiation requires $n_e$~multiplications, each being decomposed into $2 n$~controlled additions and $1$~controlled swap between two registers, giving to a total number of $2 n_e n$ ($n_e$) controlled additions (swaps between registers respectively).
Each addition needs to be modular, which can be obtained with a specific representation and a standard adder circuit.

\subsection{Coset representation}
The basic idea of the coset representation for adding $2^j \gamma$ to a quantum register encoding the integer $z$ is to extend the register for $z$ with $m$~additional qubits and to encode it into the state $\frac{1}{\sqrt{2^{m}}} \sum\limits_{k=0}^{2^{m}-1} \ket{z + k N}$.
Except at the bounds, this state is invariant under the addition of $N$.
This implies:
\begin{multline*}
	\frac{1}{\sqrt{2^{m}}} \sum\limits_{k=0}^{2^{m}-1} \ket{\left(z + 2^j \gamma + k N\right) \mod 2^{n+m}} \\
		\approx \frac{1}{\sqrt{2^{m}}} \sum\limits_{k=0}^{2^{m}-1} \ket{\left(z + 2^j \gamma \mod N\right) + k N}
\end{multline*}
\latin{i.e.\@} the modular addition of $2^j \gamma$ in the register of $z$ can be performed with a standard adder, at the cost of a small error which is exponentially suppressed when increasing $m$~\cite{Gidney2019Approximateencodedpermutations}.
Note that the resource needed to initialize the register is negligible with respect to the resource taken to implement the adder, see \autoref{appendix:coset_representation}.
Taking into account the increase in register size, this means that $2 n_e (n+m)$~controlled additions and $n_e$~controlled register swaps are needed for realizing Ekerå and Håstad's algorithm.

\subsection{Controlled operations}
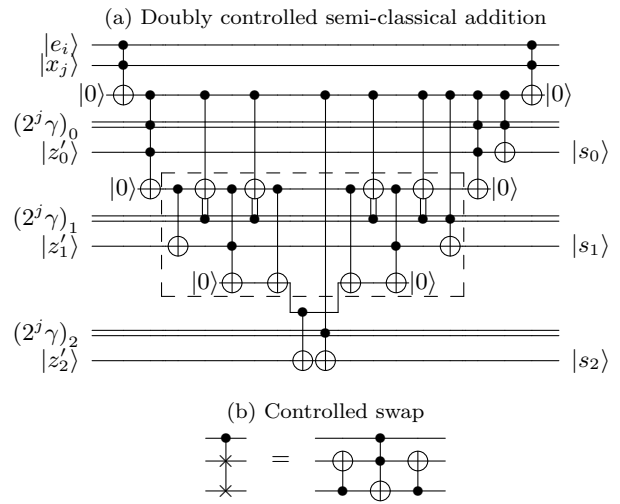
\begin{figure}[h]
\subfloat[Doubly controlled semi-classical addition]{\label{subfig:adder_semiclassical_controlled:adder_semiclassical_controlled}
	\mbox{\small \Qcircuit @R=0.6em @C=.1em {
  \lstick{\ket{e_i}}        &\ctrl{1}      &\qw      &\qw&\qw&\qw     &\qw           &\qw     &\qw       &\qw     &\qw &\qw     &\qw      &\qw    &\qw     &\qw       &\qw     &\qw              &\qw      &\qw&\qw&\qw      &\qw              &\ctrl{1}&\qw                  \\
  \lstick{\ket{x_j}}        &\ctrl{1}      &\qw      &\qw&\qw&\qw     &\qw           &\qw     &\qw       &\qw     &\qw &\qw     &\qw      &\qw    &\qw     &\qw       &\qw     &\qw              &\qw      &\qw&\qw&\qw      &\qw              &\ctrl{1}&\qw                  \\
              \push{\ket{0}}&\targ         &\ctrl{2} &\qw&\qw&\qw     &\ctrl{3}      &\qw     &\ctrl{3}  &\qw     &\qw &\qw     &\ctrl{9} &\qw    &\qw     &\ctrl{3}  &\qw     &\ctrl{3}         &\ctrl{5} &\qw&\qw&\ctrl{2} &\ctrl{2}         &\targ   &\push{\ket{0}}\qw    \\
  \lstick{{(2^j \gamma)}_0} &\cw           &\cctrl{0}&\cw&\cw&\cw     &\cw           &\cw     &\cw       &\cw     &\cw &\cw     &\cw      &\cw    &\cw     &\cw       &\cw     &\cw              &\cw      &\cw&\cw&\cctrl{0}&\cctrl{0}        &\cw     &\cw                  \\
  \lstick{\ket{z'_0}}       &\qw           &\ctrl{1} &\qw&\qw&\qw     &\qw           &\qw     &\qw       &\qw     &\qw &\qw     &\qw      &\qw    &\qw     &\qw       &\qw     &\qw              &\qw      &\qw&\qw&\ctrl{1} &\targ            &\qw     &\rstick{\ket{s_0}}\qw\\
                            &\push{\ket{0}}&\targ    &\qw&\qw&\ctrl{2}&\targ         &\ctrl{2}&\targ     &\ctrl{3}&\qw &\qw     &\qw      &\qw    &\ctrl{3}&\targ     &\ctrl{2}&\targ            &\qw      &\qw&\qw&\targ    &\push{\ket{0}}\qw&        &                     \\
  \lstick{{(2^j \gamma)}_1} &\cw           &\cw      &\cw&\cw&\cw     &\cctrl{-1}    &\cw     &\cctrl{-1}&\cw     &\cw &\cw     &\cw      &\cw    &\cw     &\cctrl{-1}&\cw     &\cctrl{-1}       &\cctrl{0}&\cw&\cw&\cw      &\cw              &\cw     &\cw                  \\
  \lstick{\ket{z'_1}}       &\qw           &\qw      &\qw&\qw&\targ   &\qw           &\ctrl{1}&\qw       &\qw     &\qw &\qw     &\qw      &\qw    &\qw     &\qw       &\ctrl{1}&\qw              &\targ    &\qw&\qw&\qw      &\qw              &\qw     &\rstick{\ket{s_1}}\qw\\
                            &              &         &   &   &        &\push{\ket{0}}&\targ   &\qw       &\targ   &\qw &        &         &       &\targ   &\qw       &\targ   &\push{\ket{0}}\qw&         &   &   &         &                 &        &                     \\
                            &              &         &   &   &        &              &        &          &        &\qwx&\ctrl{2}&\qw      &\qwx\qw&        &          &        &                 &         &   &   &         &                 &        &                     \\
  \lstick{{(2^j \gamma)}_2} &\cw           &\cw      &\cw&\cw&\cw     &\cw           &\cw     &\cw       &\cw     &\cw &\cw     &\cctrl{0}&\cw    &\cw     &\cw       &\cw     &\cw              &\cw      &\cw&\cw&\cw      &\cw              &\cw     &\cw                  \\
  \lstick{\ket{z'_2}}       &\qw           &\qw      &\qw&\qw&\qw     &\qw           &\qw     &\qw       &\qw     &\qw &\targ   &\targ    &\qw    &\qw     &\qw       &\qw     &\qw              &\qw      &\qw&\qw&\qw      &\qw              &\qw     &\rstick{\ket{s_2}}\qw
  \gategroup{6}{6}{9}{19}{1.1em}{--}
}
 }
}

\subfloat[Controlled swap]{\label{subfig:adder_semiclassical_controlled:fredkin}
	\mbox{\small \Qcircuit @R=0.4em @C=0.7em {
	& \ctrl{1}    & \qw &&   &&& \qw       & \ctrl{1} &  \qw       & \qw \\
	& \qswap      & \qw && = &&& \targ     & \ctrl{1} &  \targ     & \qw \\
	& \qswap \qwx & \qw &&   &&& \ctrl{-1} & \targ    &  \ctrl{-1} & \qw \\
}
 }
}
	\caption{
		Controlled operations.
		\protect\subref{subfig:adder_semiclassical_controlled:adder_semiclassical_controlled}:
		semi-classical adder taking each bit of the classical value $2^j \gamma = \sum_{k=0}^2 2^k {(2^j \gamma)}_k$ and the three qubits register encoding $z'$ as inputs and returning ${(2^j \gamma)}_k$ and $\ket{s_k}=\ket{{(z' + 2^j \gamma \cdot e_i \cdot x_j)}_k}$.
		The block in the dashed box uses in average $5.5$~CNOTs and $2$~Toffoli gates.
		\protect\subref{subfig:adder_semiclassical_controlled:fredkin}:
		Fredkin gate implemented with a Toffoli and two CNOT gates.
		The controlled swap between registers (as required in \autoref{fig:multiplication}) is obtained by applying it to each pair of qubits.
		}\label{fig:adder_semiclassical_controlled}
\end{figure}

A controlled swap operation between two qubits (Fredkin gate) can be performed using two CNOTs and one Toffoli gates, see \autoref{subfig:adder_semiclassical_controlled:fredkin}.
Hence the total cost for controlled swaps operating on two registers prepared in the coset representation of integers (encoded each with $n+m$~qubits) is of $2(n+m)$~CNOTs and $n+m$~Toffoli gates.

For the controlled addition, note first that since we use the coset representation of integers, a circuit for controlled addition modulo a power of two is sufficient to implement a controlled modular addition.
Such an addition can be implemented with the semi-classical adder presented in \autoref{subfig:adder_semiclassical_controlled:adder_semiclassical_controlled}, which is inspired by Refs.\,\cite{Moulton2004newquantumripple, GidneyQ2018Halvingcostquantum, Babbush2020CompilationFaultTolerant}.
It shows the basic circuit taking a classical value $2^j \gamma$ and a register encoding $z'$ and returning $2^j \gamma$ and $z' + 2^j \gamma$ if the two controlled qubits $\ket{e_i}$ and $\ket{x_j}$ are both in state $\ket{1}$.
When such an addition is applied on a quantum register encoding $z'$ using $n+m$~qubits, the block in the dashed box of \autoref{subfig:adder_semiclassical_controlled:adder_semiclassical_controlled} is repeated $n+m-2$ times, giving a mean cost of $5.5(n+m)-9$~CNOTs and $2(n+m)-1$~Toffoli gates.

\subsection{Number of gates}
Given the number of gates in the controlled addition and swap operations, the number of additions and swaps in the multiplication and the number of multiplications in the modular exponentiation, we estimate that factorization takes at leading order $11 n_e {(n+m)}^{2}$~CNOTs and $4 n_e {(n+m)}^2$~Toffoli gates.

\section{Coset representation}\label{appendix:coset_representation}

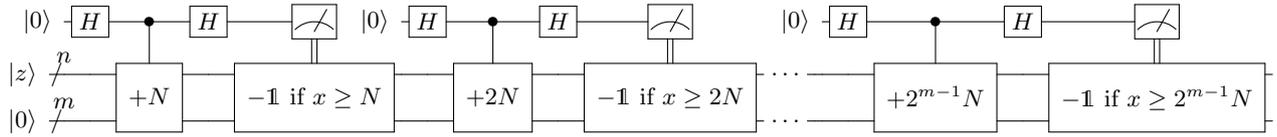
\begin{figure*}
\mbox{\small \Qcircuit @R=1em @C=0.3em {
                &       &\lstick{\ket{0}}&\gate{H}&\ctrl{1}         &\gate{H}&\meter                                            &\lstick{\ket{0}}&\gate{H}&\ctrl{1}          &\gate{H}&\meter                                             &   &            &&   &\lstick{\ket{0}}&\gate{H}&\ctrl{1}                 &\gate{H}&\meter                                                   &   \\
\lstick{\ket{z}}&\qw {/}&\ustick{n}\qw   &\qw     &\multigate{1}{+N}&     \qw&\multigate{1}{-\identite \text{ if } x \geq N}\cwx&\qw             &\qw     &\multigate{1}{+2N}&     \qw&\multigate{1}{-\identite \text{ if } x \geq 2N}\cwx&\qw&\push{\dots}&&\qw&\qw             &\qw     &\multigate{1}{+2^{m-1} N}&     \qw&\multigate{1}{-\identite \text{ if } x \geq 2^{m-1}N}\cwx&\qw\\
\lstick{\ket{0}}&\qw {/}&\ustick{m}\qw   &\qw     &       \ghost{+N}&     \qw&       \ghost{-\identite \text{ if } x \geq N}    &\qw             &\qw     &       \ghost{+2N}&     \qw&       \ghost{-\identite \text{ if } x \geq 2N}    &\qw&\push{\dots}&&\qw&\qw             &\qw     &       \ghost{+2^{m-1} N}&     \qw&       \ghost{-\identite \text{ if } x \geq 2^{m-1}N}    &\qw\\
}
 }
\caption{Preparation proposed in~\cite[Fig.\,1]{Gidney2019Approximateencodedpermutations} of a quantum register with $n+m$~qubits in the state $\frac{1}{\sqrt{2^{m}}} \sum\limits_{k=0}^{2^{m}-1} \ket{z + k N}$ as requested in the initialization of the coset representation.
The first controlled operation adds the integer $N$ to the register made with $n+m$~qubits provided that the ancillary qubit is in state $\ket{1}$.
The first classically controlled operation aims to change the phase of the input state encoded in $n+m$~qubits if and only if the result of the measurement is $1$ and the number encoded in the $n+m$~qubits is larger or equal than $N$.
In case one of the two conditions is not met, the input state is unchanged.}\label{fig:coset_init}
\end{figure*}

Modular addition is typically implemented with variants of the addition: an addition, a comparison, a controlled correction and the clean-up of ancillary qubits~\cite{EkertPRA1996Quantumnetworkselementary}.
As exposed in main text, coset representation of integers, introduced by Zalka~\cite{Zalka2006Shorsalgorithmfewer} and formalized by Gidney~\cite{Gidney2019Approximateencodedpermutations}, can be used to approximate the modular addition with a single standard adder circuit.

The basic idea of the coset representation for adding $2^j \gamma$ modulo $N$ to a quantum register encoding the integer $z$ is to extend the register for $z$ with $m$~additional qubits and to encode it into the state $\frac{1}{\sqrt{2^{m}}} \sum\limits_{k=0}^{2^{m}-1} \ket{z + k N}$.
Except at the bounds, this state is invariant under the addition of $N$.
This implies
\begin{multline}\label{eq:coset}
	\frac{1}{\sqrt{2^{m}}} \sum\limits_{k=0}^{2^{m}-1} \ket{\left(z + 2^j \gamma + k N\right) \mod 2^{n+m}} \\
		\approx \frac{1}{\sqrt{2^{m}}} \sum\limits_{k=0}^{2^{m}-1} \ket{\left(z + 2^j \gamma \mod N\right) + k N}
\end{multline}
\latin{i.e.\@} the modular addition of $2^j \gamma$ in the register of $z$ can be performed with a standard adder (modulo $2^{n+m}$), at the cost of a small error which is exponentially suppressed when increasing $m$~\cite{Gidney2019Approximateencodedpermutations}.
Note also that the precision is improved if instead of adding $2^j \gamma$, one adds $2^j \gamma \mod N$ (which does not change the result of the sum since we consider the sum modulo $N$).
This is possible each time the quantity to add is known classically.

The goal of the first subsection is to show that the resource needed to extend the register encoding $\ket{z}$ into the state $\frac{1}{\sqrt{2^{m}}} \sum\limits_{k=0}^{2^{m}-1} \ket{z + k N}$, as requested in this representation, is negligible with respect to the resource taken to implement the modular exponentiation.
In the second subsection, we show that the coset representation is compatible with the modular multiplication circuit presented in the main text.

In the two next subsections, the coset representation is considered for additions modulo $N$; $n$ is the number of bits encoding $N$, and $m$ the number of qubits added to the register for the coset representation.

\subsection{Initialization}
\begin{figure}[h]
\subfloat[Semi-classical comparison and correction]{\label{subfig:comparison:comparison}
	\mbox{\small \Qcircuit @R=0.7em @C=0.2em {
	\lstick{x_0} &\ctrl{2} &\qw     &\qw       &\qw     &\qw       &\qw     &\qw &\qw     &\qw       &\qw     &\qw       &\qw     &\qw     &\qw     &\qw       &\qw     &\qw       &\qw     &\qw    &\qw     &\qw       &\qw     &\qw       &\qw     &\ctrl{2} &\rstick{x_0} \qw\\
	\lstick{y'_0}&\cctrl{0}&\cw     &\cw       &\cw     &\cw       &\cw     &\cw &\cw     &\cw       &\cw     &\cw       &\cw     &\cw     &\cw     &\cw       &\cw     &\cw       &\cw     &\cw    &\cw     &\cw       &\cw     &\cw       &\cw     &\cctrl{0}&\rstick{y'_0}\cw\\
	             &         &\ctrl{1}&\targ     &\ctrl{1}&\targ     &\ctrl{3}&\qw &\qw     &\qw       &\qw     &\qw       &\qw     &\qw     &\qw     &\qw       &\qw     &\qw       &\qw     &\qw    &\ctrl{3}&\targ     &\ctrl{1}&\targ     &\ctrl{1}&\qw      &                \\
	\lstick{x_1} &\qw      &\targ   &\qw       &\ctrl{2}&\qw       &\qw     &\qw &\qw     &\qw       &\qw     &\qw       &\qw     &\qw     &\qw     &\qw       &\qw     &\qw       &\qw     &\qw    &\qw     &\qw       &\ctrl{2}&\qw       &\targ   &\qw      &\rstick{x_1} \qw\\
	\lstick{y'_1}&\cw      &\cw     &\cctrl{-2}&\cw     &\cctrl{-2}&\cw     &\cw &\cw     &\cw       &\cw     &\cw       &\cw     &\cw     &\cw     &\cw       &\cw     &\cw       &\cw     &\cw    &\cw     &\cctrl{-2}&\cw     &\cctrl{-2}&\cw     &\cw      &\rstick{y'_1}\cw \gategroup{3}{3}{6}{25}{1em}{--}\\
	             &         &        &          &        &\qw       &\targ   &\qw &        &          &        &          &        &        &        &          &        &          &        &       &\targ   &\qw       &\qw     &          &        &         &                \\
	             &         &        &          &        &          &        &\qwx&\ctrl{1}&\targ     &\ctrl{1}&\targ     &\ctrl{3}&\qw     &\ctrl{3}&\targ     &\ctrl{1}&\targ     &\ctrl{1}&\qw\qwx&        &          &        &          &        &         &                \\
	\lstick{x_2} &\qw      &\qw     &\qw       &\qw     &\qw       &\qw     &\qw &\targ   &\qw       &\ctrl{2}&\qw       &\qw     &\qw     &\qw     &\qw       &\ctrl{2}&\qw       &\targ   &\qw    &\qw     &\qw       &\qw     &\qw       &\qw     &\qw      &\rstick{x_2} \qw\\
	\lstick{y'_2}&\cw      &\cw     &\cw       &\cw     &\cw       &\cw     &\cw &\cw     &\cctrl{-2}&\cw     &\cctrl{-2}&\cw     &\cw     &\cw     &\cctrl{-2}&\cw     &\cctrl{-2}&\cw     &\cw    &\cw     &\cw       &\cw     &\cw       &\cw     &\cw      &\rstick{y'_2}\cw \\
	             &         &        &          &        &          &        &    &        &          &        &\qw       &\targ   &\gate{Z}&\targ   &\qw       &\qw     &          &        &       &        &          &        &          &        &         &                \\
} }
}

\subfloat[AND computation]{\label{subfig:comparison:and}
	\hspace{0.2em}
	\mbox{\small \Qcircuit @R=0.5em @C=0.5em {
	&&\ctrl{1}&\qw && &&&\qw           &\ctrl{1}&\qw\\
	&&\ctrl{1}&\qw &&=&&&\qw           &\ctrl{1}&\qw\\
	&&        &\qw && &&&\push{\ket{0}}&\targ   &\qw\\
}
 }
	\hspace{0.2em}
}
\hfil
\subfloat[AND uncomputation]{\label{subfig:comparison:deand}
	\hspace{0.4em}
	\mbox{\small \Qcircuit @R=0.5em @C=0.5em {
	&&\ctrl{1}&\qw && &&&\qw  &\ctrl{1}&\qw               &\qw\\
	&&\ctrl{1}&\qw &&=&&&\qw  &\ctrl{1}&\qw               &\qw\\
	&&\qw     &    && &&&\qw  &\targ   &\push{\ket{0}} \qw&   \\
}
 }
	\hspace{0.4em}
}
\caption{
    \protect\subref{subfig:comparison:comparison}:
	Circuit inspired from~\cite[Fig.\,17]{Babbush2020CompilationFaultTolerant} which compares the integer $x$ encoded in $n+m$~qubits and the integer $y<2^{n+m}$ known classically, and returns $-\ket{x}$ if and only if $x \geq y$.
	This is done in three steps: i) compute the carries of $y'+x$ with $y' = 2^{n+m} - y$, ii) apply a $Z$ operation on the last carry and iii) uncompute the carries.\\
	\protect\subref{subfig:comparison:and} and \protect\subref{subfig:comparison:deand}: circuits defining the notations used to compute and uncompute an AND operation, as introduced in~\cite{GidneyQ2018Halvingcostquantum,NevenPRX2018EncodingElectronicSpectra} where the authors give efficient implementations in terms of $T$ (or $\frac{\pi}{4}$) gates.
	When only one quantum control appear, it uses a CNOT instead of a Toffoli gate, and it can be removed by directly using the control bit instead of the ancillary.
	}\label{fig:comparison}
\end{figure}
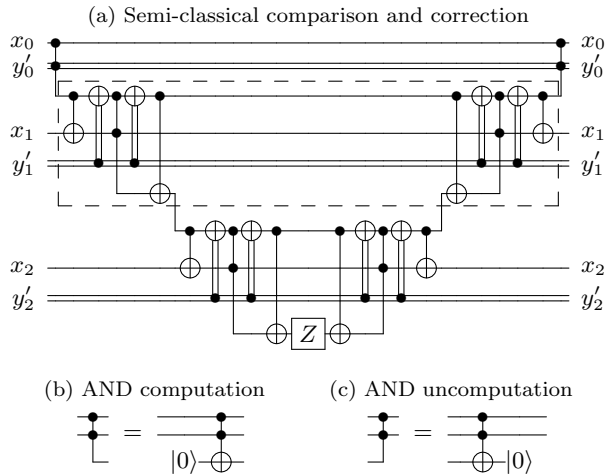

Starting from a register with $n$~qubits in state $\ket{z}$, the initialization of the coset representation consists in preparing the state $\frac{1}{\sqrt{2^{m}}} \sum\limits_{k=0}^{2^{m}-1} \ket{z + k N}$ in an extended register of size $n+m$.
This is done by performing successive additions, each controlled by an ancillary qubit prepared in the state $\frac{\ket{0}+\ket{1}}{\sqrt{2}}$ ($m$~ancillary qubits in total) which is then uncomputed, see \autoref{fig:coset_init}.
The controlled addition is performed using the circuit presented in \autoref{subfig:adder_semiclassical_controlled:adder_semiclassical_controlled} with only one control qubit.
The uncomputation of the ancillary qubit is based on a measurement and depending on the result, a conditioned correction is realized, see \autoref{fig:coset_init}.
Let us detail the uncomputation of the first ancillary qubit presented in \autoref{fig:coset_init}.
When the result of the measurement is $0$, the register made with $n+m$~qubits is projected into $\frac{1}{\sqrt{2}}\left(\ket{z} + \ket{z+N}\right)$.
When the result is $1$, the register state is $\frac{1}{\sqrt{2}}\left(\ket{z} - \ket{z+N}\right)$ and the operation $-\identite$ needs to be applied to the component $\ket{z+N}$, \latin{i.e.\@}, when the state of the register encodes an integer larger than $N$.
In order to implement the conditioned operations for decomputing the $m$~ancillary qubits, we need to compare the value $x$ encoded in the quantum register of size $n+m$ and an integer $y$ known classically satisfying $0 < y \leq 2^{m-1}N < 2^{n+m}$ (see the last umcomputation in \autoref{fig:coset_init}) \latin{i.e.\@} that can be written with $n+m$~bits.
This comparison is implemented using the circuit presented in \autoref{subfig:comparison:comparison}.
First, the value $2^{n+m}-y = y'$ is computed classically.
Then the last carry of the sum of $x$ and $y'$ is computed with a circuit derived from the addition.
If the value of this carry is $1$, we conclude that $x \geq y$, otherwise $x < y$.
A $Z$ gate is thus applied on the qubit encoding the last carry, before uncomputing the carries.
The register ends up in state $\pm\ket{x}$ depending on the relative value between $x$ and $y$, as desired.

Each controlled addition and correction costs $\bigO{n+m}$ gates.
This operation is repeated $m$~times, giving a total cost of the coset representation initialization of the order $\bigO{m(n+m)}$.

In the modular exponentiation algorithm, the two registers at the bottom of \autoref{fig:multiplication} need to be prepared initially in $x=1$ and $0$ respectively.
Initializing them in the coset representation $\frac{1}{\sqrt{2^{m}}} \sum\limits_{k=0}^{2^{m}-1} \ket{1 + k N}$ and $\frac{1}{\sqrt{2^{m}}} \sum\limits_{k=0}^{2^{m}-1} \ket{k N}$ takes $\bigO{m(n+m)}$~gates which is negligible compared to the cubic cost of the full exponentiation.
Note however, that the cost of this initialization is taken into account in our script for the evaluation of the whole algorithm cost.

\begin{figure*}
\subfloat[Multiplication for the exponentiation, with $i=2$ and $w_e=2$, decomposed into product-additions.]{\label{subfig:exponentiation:multiplication}
	\mbox{\small \providecommand{\multicontrolgate}[2]{\multimeasure{#1}{#2}}
\providecommand{\controlgate}[1]{\measure{#1}}
\Qcircuit @R=0.5em @C=0.5em {
	                &\qw   &\qw                                           &\qw &&& && &\qw             &\qw   &\qw                                           &\qw                                                            &\qw&\qw          &\qw                                           \\
	                &\qw   &\qw                                           &\qw &&& && &\qw             &\qw   &\qw                                           &\qw                                                            &\qw&\qw          &\qw                                           \\
	                &\qw   &\multicontrolgate{1}{\text{Input }e_{i:i+w_e}}&\qw &&& && &\qw             &\qw   &\multicontrolgate{1}{\text{Input }e_{i:i+w_e}}&\multicontrolgate{1}{\text{Input }e_{i:i+w_e}}                 &\qw&\qw          &\qw                                           \\
	                &\qw   &              \ghost{\text{Input }e_{i:i+w_e}}&\qw &&&=&& &\qw             &\qw   &              \ghost{\text{Input }e_{i:i+w_e}}&              \ghost{\text{Input }e_{i:i+w_e}}                 &\qw&\qw          &\qw                                           \\
	                &\qw   &\qw\qwx                                       &\qw &&& && &\qw             &\qw   &\qw\qwx                                       &\qw\qwx                                                        &\qw&\qw          &\qw                                           \\
	                &\qw   &\qw\qwx                                       &\qw &&& && &\qw             &\qw   &\qw\qwx                                       &\qw\qwx                                                        &\qw&\qw          &\qw \inputgroupv{1}{6}{0.75em}{2.2em}{\ket{e}}\\
	                &      &\qwx                                          &    &&& && &                &      &\qwx                                          &\qwx                                                           &   &             &                                              \\
	                &      &\qwx                                          &    &&& && &                &      &\qwx                                          &\qwx                                                           &   &             &                                              \\
	\lstick{\ket{x}}&{/}\qw&\gate{\times g^{2^i e_{i:i+w_e}} \mod{N}}\qwx &\qw &&& && &\lstick{\ket{x}}&{/}\qw&\controlgate{\text{Input }x}\qwx              &\gate{+ \bar{x} \left(-g^{-2^i e_{i:i+w_e}}\right) \mod{N}}\qwx&\qw&\link{1}{-1} &\qw                                           \\
	                &      &                                              &    &&& && &\lstick{\ket{0}}&{/}\qw&\gate{+ x g^{2^i e_{i:i+w_e}} \mod{N}}\qwx    &\controlgate{\text{Input }\bar{x}}\qwx                         &\qw&\link{-1}{-1}&\push{\bra{0}}\qw                             \\
}
 }
}

\medskip

\subfloat[Windowed product-addition, as needed in \protect\subref{subfig:exponentiation:multiplication}, with the windows size $w_m = 3$.]{\label{subfig:exponentiation:product_add}
	\mbox{\small \providecommand{\multicontrolgate}[2]{\multimeasure{#1}{#2}}
\providecommand{\controlgate}[1]{\measure{#1}}
\Qcircuit @R=0.5em @C=0.5em {
	&{/}\qw&\ustick{w_e}\qw&\controlgate{\text{Input }e_{i:i+w_e}}    &\qw &&& && &{/}\qw&\ustick{w_e}\qw&\controlgate{\text{Input }e_{i:i+w_e}}                &\controlgate{\text{Input }e_{i:i+w_e}}                 &\qw&\\
	&\qw   &\qw            &\multicontrolgate{5}{\text{Input }x}\qwx  &\qw &&& && &\qw   &\qw            &\multicontrolgate{2}{\text{Input }x_{0:3}}\qwx        &\qw\qwx                                                &\qw&\\
	&\qw   &\qw            &              \ghost{\text{Input }x}      &\qw &&& && &\qw   &\qw            &              \ghost{\text{Input }x_{0:3}}            &\qw\qwx                                                &\qw&\\
	&\qw   &\qw            &              \ghost{\text{Input }x}      &\qw &&& && &\qw   &\qw            &              \ghost{\text{Input }x_{0:3}}            &\qw\qwx                                                &\qw&\\
	&\qw   &\qw            &              \ghost{\text{Input }x}      &\qw &&&=&& &\qw   &\qw            &\qw\qwx                                               &\multicontrolgate{2}{\text{Input }x_{3:6}}\qwx         &\qw&\\
	&\qw   &\qw            &              \ghost{\text{Input }x}      &\qw &&& && &\qw   &\qw            &\qw\qwx                                               &              \ghost{\text{Input }x_{3:6}}             &\qw&\\
	&\qw   &\qw            &              \ghost{\text{Input }x}      &\qw &&& && &\qw   &\qw            &\qw\qwx                                               &              \ghost{\text{Input }x_{3:6}}             &\qw&\\
	&{/}\qw&\qw            &\gate{+ x g^{2^i e_{i:i+w_e}} \mod{N}}\qwx&\qw &&& && &{/}\qw&\qw            &\gate{+ 2^{0} x_{0:3} g^{2^i e_{i:i+w_e}} \mod{N}}\qwx& \gate{+ 2^{3} x_{3:6} g^{2^i e_{i:i+w_e}} \mod{N}}\qwx&\qw&\\
}
 }
}

\medskip

\subfloat[Modular addition of a number read with a table lookup, as needed in \protect\subref{subfig:exponentiation:product_add}.]{\label{subfig:exponentiation:lookup_add}
	\mbox{\small \providecommand{\controlgate}[1]{\measure{#1}}
\Qcircuit @R=1em @C=0.75em {
	&{/}\qw&\ustick{\hspace{-0.5em}w_e+w_m}\qw&\controlgate{\text{Input }k}&\qw && && &{/}\qw          &\ustick{\hspace{-0.5em}w_e+w_m}\qw&\qw          &\controlgate{\text{Input }k}&\qw&\qw                  &\qw                            &\controlgate{\text{Input }k}&\qw                &\\
	&      &                                  &                    \qwx    &    &&=&& &\lstick{\ket{0}}&{/}\qw                            &\ustick{n}\qw&\gate{\text{Load }T_k}\qwx  &\qw&\ustick{\ket{T_k}}\qw&\controlgate{\text{Input } T_k}&\gate{\text{Unload }T_k}\qwx&\rstick{\ket{0}}\qw&\\
	&{/}\qw&\qw                               &\gate{+ T_k \mod{N}}\qwx    &\qw && && &{/}\qw          &\qw                               &\qw          &\qw                         &\qw&\qw                  &\gate{+T_k \mod N}\qwx         &\qw                         &\qw                &\\
}
 }
}

\caption{Windowed arithmetic subcircuits for the modular exponentiation.
When not specified, the register size is $n+m$~qubits (register encoded into the coset representation of integers).}\label{fig:exponentiation}
\end{figure*}
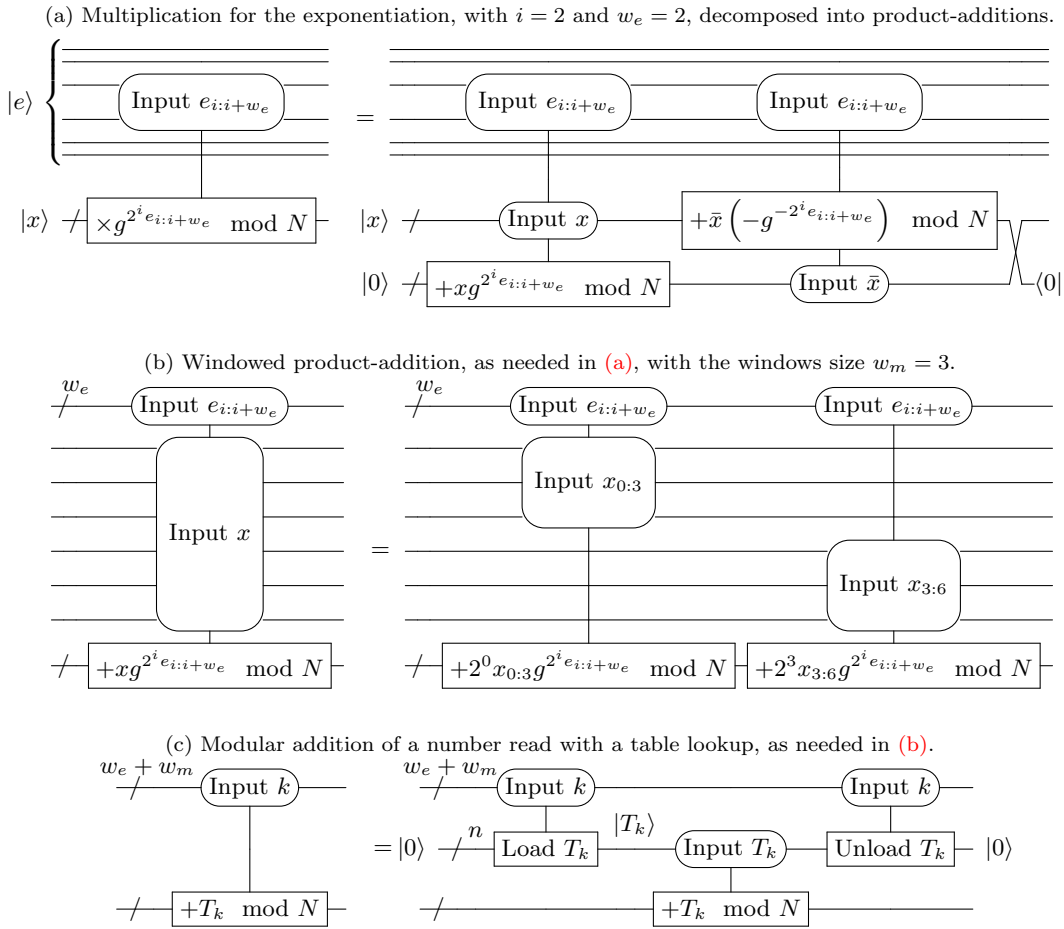

\subsection{Compatibility with the multiplication}
When computing the multiplications from sequences of two product-additions (see \autoref{fig:multiplication} of main text), the input register encoding $x$ and the ancillary register are used both as control and target of the product-additions.
We here check that having the control register encoded in the coset representation is not a problem for performing the multiplication.

Let us consider the first product-addition used to implement the multiplication shown in the bottom part of \autoref{fig:multiplication}.
In the coset representation, the input $x$ and ancillary registers are in the state $\frac{1}{2^{m}} \sum\limits_{k=0}^{2^m-1} \ket{x + k N} \sum\limits_{k'=0}^{2^m-1} \ket{0 + k' N}$ meaning that after the product-addiction, their state ends up in $\frac{1}{2^{m}} \sum\limits_{k=0}^{2^m-1} \sum\limits_{k'=0}^{2^m-1} \ket{x + k N} \ket{(x + k N)g^{2^i} + k' N \mod{2^{n+m}}}$.
As $kN g^{2^i} + k'N$ is a multiple of $N$, the obtained state is very close to $\frac{1}{2^{m}} \sum\limits_{k=0}^{2^m-1} \ket{x + k N} \sum\limits_{k'=0}^{2^m-1} \ket{x g^{2^i} + k' N}$  thanks to the coset representation itself, \latin{cf.\@} \autoref{eq:coset}.
The latter corresponds to the desired state.

\section{Windowed arithmetic}\label{appendix:windowed_arithmetic}
In order to reduce the number of multiplications and additions in the exponentiation algorithm, we use windowed arithmetic circuits~\cite{Gidney2019Windowedquantumarithmetic}.
They consist in grouping the bits of $e$ for controlling each multiplication, hence reducing the number of multiplications.
Similarly, for each multiplication, the input bits are grouped to reduce the number of additions composing each multiplication.

The next subsection shows the details of the decomposition of the exponentiation into elementary additions of the form $+T_k \mod N$ where the quantity $T_k$ depends on the value of an integer $k$.
These specific additions are implemented in three steps, that are presented in separated subsequent subsections.

\subsection{Windowed exponentiation and multiplication}
Let us start by specifying the notations.
We label the binary form of $e$ as
\begin{equation}
e_{n_e-1} \ldots e_{i+w_e} \overbrace{e_{i+w_e-1} \ldots e_{i}}^{e_{i:i+w_e}} \ldots e_{2} e_{1} e_{0}
\end{equation}
\latin{i.e.\@} $e_j$ is the $j$th bit of $e$.
Let also $e_{i:i+w_e}$ be defined as
\begin{equation}
e_{i:i+w_e} = \sum\limits_{j=i}^{i+w_e-1} 2^{j-i} e_j
\end{equation}
\latin{i.e.\@} $e_{i:i+w_e}$ is the number whose bit decomposition is given by the bits of $e$ starting at index $i$ and taking $w_e$~bits.
The strategy for computing the exponentiation using windowed arithmetic consists in decomposing exponent $e$ in terms of numbers $e_{i:i+w_e}$
\begin{equation}
e = \sum\limits_{\mathclap{\substack{0 \leq i < n_e \\ i \equiv 0 \mod w_e}}} 2^{i} e_{i:i+w_e},
\end{equation}
such that
\begin{equation}\label{eq:decompose_exponent_windowarithmetic}
g^e
	= \prod\limits_{\mathclap{\substack{0 \leq i < n_e \\ i \equiv 0 \mod w_e}}} g^{2^{i} e_{i:i+w_e}}.
\end{equation}
The comparison with the decomposition of $g^e$ presented in \autoref{eq:decompose_exponent} clearly shows that windowed exponentiation divides the number of multiplications by $w_e$.

As for the standard algorithm, the multiplications of the product~\eqref{eq:decompose_exponent_windowarithmetic} are implemented successively and each multiplication is decomposed into a sequence of two product-additions, as shown in \autoref{subfig:exponentiation:multiplication}.
The difference is that the added number now depends on the number $e_{i:i+w_e}$.

The product-addition is also performed in a windowed way~\cite{Gidney2019Windowedquantumarithmetic}.
\autoref{subfig:exponentiation:product_add} shows in particular how the first product-addition needed for each multiplication is performed using windows for input $x$ of size $w_m=3$.

\autoref{subfig:exponentiation:lookup_add} finally shows the implementation of an addition $+T_k \mod N$ of a quantity $T_k$ that depends on the value $k$.
It requires three steps.
First, the number $T_k$ is loaded into an ancillary register.
Second, this number is unconditionally added to the desired register and finally the ancillary register is cleaned up.
Note that the value of $T_k$ (given by $T_{k_1, k_2} = 2^{i} k_1 g^{2^i k_2}$, with $k_1 = e_{i:i+w_e}$ and $k_2 = x_{i:i+w_m}$, $k$ being the concatenation of $k_1$ and $k_2$) to be added is known classically.
Its addition being realized modulo $N$, its value can be computed modulo $N$ before being loaded.
$n$~bits are thus sufficient to encode $T_k$.

Loading a value $T_k$ into a quantum register is done using a quantum table lookup circuit which we discuss right after.
The subsequent subsection is dedicated to the task aiming to unload the value $T_k$ and reset the register in state $\ket{0}$.
The last subsection is dedicated to the requested addition.

\subsection{Table lookup}
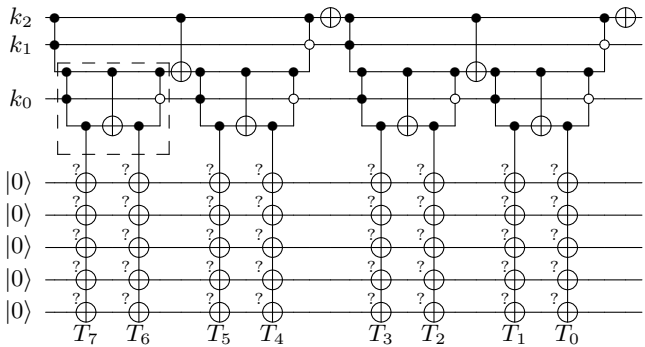
\begin{figure}[h]
	\mbox{\small \hspace{1em} 

\Qcircuit @C=0.25em @R=0.5em {
\lstick{k_2}    &\ctrl{1}&\qw     &\qw         &\qw     &\qw         &\qw      &\ctrl{2}&\qw     &\qw         &\qw     &\qw         &\qw      &\ctrl{1} &\targ&\ctrl{1}&\qw     &\qw         &\qw     &\qw         &\qw      &\ctrl{2}&\qw     &\qw         &\qw     &\qw         &\qw      &\ctrl{1} &\targ&\qw \\
\lstick{k_1}    &\ctrl{1}&\qw     &\qw         &\qw     &\qw         &\qw      &\qw     &\qw     &\qw         &\qw     &\qw         &\qw      &\ctrlo{1}&\qw  &\ctrl{1}&\qw     &\qw         &\qw     &\qw         &\qw      &\qw     &\qw     &\qw         &\qw     &\qw         &\qw      &\ctrlo{1}&\qw  &\qw \\
                &        &\ctrl{1}&\qw         &\ctrl{2}&\qw         &\ctrl{1} &\targ   &\ctrl{1}&\qw         &\ctrl{2}&\qw         &\ctrl{1} &\qw      &     &        &\ctrl{1}&\qw         &\ctrl{2}&\qw         &\ctrl{1} &\targ   &\ctrl{1}&\qw         &\ctrl{2}&\qw         &\ctrl{1} &\qw      &     &    \\
\lstick{k_0}    &\qw     &\ctrl{1}&\qw         &\qw     &\qw         &\ctrlo{1}&\qw     &\ctrl{1}&\qw         &\qw     &\qw         &\ctrlo{1}&\qw      &\qw  &\qw     &\ctrl{1}&\qw         &\qw     &\qw         &\ctrlo{1}&\qw     &\ctrl{1}&\qw         &\qw     &\qw         &\ctrlo{1}&\qw      &\qw  &\qw \\
                &        &        &\ctrl{3}    &\targ   &\ctrl{3}    &\qw      &        &        &\ctrl{3}    &\targ   &\ctrl{3}    &\qw      &         &     &        &        &\ctrl{3}    &\targ   &\ctrl{3}    &\qw      &        &        &\ctrl{3}    &\targ   &\ctrl{3}    &\qw      &         &     &    \\
                &        &        &            &        &            &         &        &        &            &        &            &         &         &     &        &        &            &        &            &         &        &        &            &        &            &         &         &     &    \\
                &        &        &            &        &            &         &        &        &            &        &            &         &         &     &        &        &            &        &            &         &        &        &            &        &            &         &         &     &    \\
\lstick{\ket{0}}&\qw     &\qw     &\targ_?     &\qw     &\targ_?     &\qw      &\qw     &\qw     &\targ_?     &\qw     &\targ_?     &\qw      &\qw      &\qw  &\qw     &\qw     &\targ_?     &\qw     &\targ_?     &\qw      &\qw     &\qw     &\targ_?     & \qw    &\targ_?     &\qw      &\qw      &\qw  &\qw \\
\lstick{\ket{0}}&\qw     &\qw     &\targ_? \qwx&\qw     &\targ_? \qwx&\qw      &\qw     &\qw     &\targ_? \qwx&\qw     &\targ_? \qwx&\qw      &\qw      &\qw  &\qw     &\qw     &\targ_? \qwx&\qw     &\targ_? \qwx&\qw      &\qw     &\qw     &\targ_? \qwx& \qw    &\targ_? \qwx&\qw      &\qw      &\qw  &\qw \\
\lstick{\ket{0}}&\qw     &\qw     &\targ_? \qwx&\qw     &\targ_? \qwx&\qw      &\qw     &\qw     &\targ_? \qwx&\qw     &\targ_? \qwx&\qw      &\qw      &\qw  &\qw     &\qw     &\targ_? \qwx&\qw     &\targ_? \qwx&\qw      &\qw     &\qw     &\targ_? \qwx& \qw    &\targ_? \qwx&\qw      &\qw      &\qw  &\qw \\
\lstick{\ket{0}}&\qw     &\qw     &\targ_? \qwx&\qw     &\targ_? \qwx&\qw      &\qw     &\qw     &\targ_? \qwx&\qw     &\targ_? \qwx&\qw      &\qw      &\qw  &\qw     &\qw     &\targ_? \qwx&\qw     &\targ_? \qwx&\qw      &\qw     &\qw     &\targ_? \qwx& \qw    &\targ_? \qwx&\qw      &\qw      &\qw  &\qw \\
\lstick{\ket{0}}&\qw     &\qw     &\targ_? \qwx&\qw     &\targ_? \qwx&\qw      &\qw     &\qw     &\targ_? \qwx&\qw     &\targ_? \qwx&\qw      &\qw      &\qw  &\qw     &\qw     &\targ_? \qwx&\qw     &\targ_? \qwx&\qw      &\qw     &\qw     &\targ_? \qwx& \qw    &\targ_? \qwx&\qw      &\qw      &\qw  &\qw \gategroup{3}{3}{6}{7}{0.5em}{--} \\
                &        &        & T_7        &        &  T_6       &         &        &        &  T_5       &        &  T_4       &         &         &     &        &        &  T_3       &        &  T_2       &         &        &        &  T_1       &        &  T_0       &         &         &     &    \\
                &        &        &            &        &            &         &        &        &            &        &            &         &         &     &        &        &            &        &            &         &        &        &            &        &            &         &         &     &    \\
} }
\caption{
	Example of a quantum table lookup.
	For a basis state $\ket{k}$ specifying the address of the number $T_k$ from a classical table, the quantum table lookup maps basis states $\ket{k}\ket{0}$ into $\ket{k}\ket{T_k}$.
	Here $k$ and the output are composed respectively of 3 and 5~qubits.
	The notations for the AND computation and uncomputation is presented in \autoref{fig:comparison}.
	Black and white circles are controls on the $\ket{1}$ and $\ket{0}$ states respectively.
	The question mark on the controlled NOT means that a controlled NOT is applied on qubit $i$ only when the $i$th bit of $T_k$ takes the value $1$.
	}\label{fig:qrom}
\end{figure}

The quantum table lookup proposed in~\cite{NevenPRX2018EncodingElectronicSpectra}, produces the following operation on basis states: $\ket{k}\ket{x} \mapsto \ket{k}\ket{x \oplus T_k}$ with $\oplus$ the bitwise XOR operator.
For state preparation, as required in the first step of the operation presented in \autoref{subfig:exponentiation:lookup_add}, the target register starts in the state $\ket{0}$ such that control and target registers end in $\ket{k}\ket{T_k}$.
The circuit presented in \autoref{fig:qrom} shows the principle of this operation with registers for $k$ and $T_k$ composed respectively of 3 and 5~qubits.

Concretely, the numbers $T_k$ specify the set of controlled NOT to be used (the question mark on the controlled NOT means that a controlled NOT is applied on qubit $i$ only when the $i$th bit of $T_k$ takes the value $1$).
The circuit operating on the bits $k_i$ of $k$ prepares the last ancillary qubit (line 5 from the top) in the state $\ket{1}$ at the time (specified by $k$) where the gates corresponding to $T_k$ are applied, and $\ket{0}$ otherwise.
The building block of the circuit is boxed in \autoref{fig:qrom}.
It uses 1 CNOT, 1 AND computation and uncomputation.
Given that $k$ is encoded into the number of bits $w_e+w_n$ and can thus take $2^{w_e+w_n}$~different values, the number of blocks in the upper part of \autoref{fig:qrom} is given by $\sum\limits_{j=1}^{w_e+w_m-1} 2^j = 2^{w_e-w_n}-2$.
This means that $2^{w_e + w_m} - 2$~CNOT gates, $2^{w_e + w_m} - 2$~AND computations and uncomputations are needed to implement these blocks.
Moreover, the number of controlled multi-NOT gates to load the value $T_k$ is given by $2^{w_e+w_m}$, each gate being decomposed into $n/2$~CNOT in average since $T_k$ takes $n$~bits.
When including the (two) NOT gates operating on the highest bit of $k$, we conclude that the table lookup uses $2$~NOT gates, $2^{w_e + w_m} - 2 + 2^{w_e+w_m-1}n$~CNOT gates, $2^{w_e + w_m} - 2$~AND computations and uncomputations (corresponding to $2 \times \left(2^{w_e + w_m} - 2\right)$~Toffoli gates).

\subsection{Table unlookup}
The purpose of the table unlookup operation (last step in \autoref{subfig:exponentiation:lookup_add}) is to map the state $\sum\limits_{k} \alpha_k \ket{k} \ket{T_k}$ into $\sum\limits_k \alpha_k \ket{k}$, where $\alpha_k$ are some complex coefficients.
A natural way to do this mapping is to apply again the lookup operation described in the previous subsection.
Since the lookup operates on the computational basis following $\ket{k}\ket{x} \mapsto \ket{k}\ket{x \oplus T_k}$ where $\oplus$ stands for the bitwise XOR operator, by linearity it maps $\sum\limits_{k} \alpha_k \ket{k} \ket{T_k} \mapsto \sum\limits_k \alpha_k \ket{k} \ket{0}$, the latter corresponding to the desired state when simply discarding the qubits previously encoding the numbers $T_k$.

However, a more efficient measurement-based technique is possible, as shown in  Ref.\,\cite[Appendix~C]{BabbushQ2019QubitizationArbitraryBasis} and improved in Ref.\,\cite{Gidney2019Windowedquantumarithmetic}.
The principle consists in starting by measuring the register encoding $T_k$ in the $X$ basis before applying a phase shift conditioned on the result of measurements.
For a more detailed explanation, let us start to expand the qubits encoding the numbers $T_k$ in bits indexed by $j$ (${(T_k)}_j$ being the $j$th bit of $T_k$).
The state before the uncomputation can be written as
\begin{equation}
    \sum\limits_k \alpha_k \ket{k} \bigotimes\limits_j \ket{{(T_k)}_j}_j.
\end{equation}
Let us now focus on a specific qubit indexed by $j^*$.
We label $\mathcal{K}_0 = \lbrace k \mid {(T_k)}_{j^*} = 0 \rbrace$ and $\mathcal{K}_1 = \lbrace k \mid {(T_k)}_{j^*} = 1 \rbrace$.
The state before the uncomputation can be rewritten as
\begin{multline}
    \left[\sum\limits_{k \in \mathcal{K}_0} \alpha_k \ket{k} \bigotimes\limits_{j \neq j^*} \ket{{(T_k)}_j}_j \right] \ket{0}_{j^*} \\
        + \left[\sum\limits_{k \in \mathcal{K}_1} \alpha_k \ket{k} \bigotimes\limits_{j \neq j^*} \ket{{(T_k)}_j}_j \right] \ket{1}_{j^*}.
\end{multline}
By applying a Hadamard gate on the $j^*$th qubit, we obtain
\begin{multline}
    \frac{1}{\sqrt{2}}\left[
    	\begin{multlined}
    	\sum\limits_{k \in \mathcal{K}_0} \alpha_k \ket{k} \bigotimes\limits_{j \neq j^*} \ket{{(T_k)}_j}_j \\
    	+ \sum\limits_{k \in \mathcal{K}_1} \alpha_k \ket{k} \bigotimes\limits_{j \neq j^*} \ket{{(T_k)}_j}_j
    	\end{multlined}
   	\right] \ket{0}_{j^*}\\
	+ \frac{1}{\sqrt{2}}\left[
		\begin{multlined}
		\sum\limits_{k \in \mathcal{K}_0} \alpha_k \ket{k} \bigotimes\limits_{j \neq j^*} \ket{{(T_k)}_j}_j \\
		- \sum\limits_{k \in \mathcal{K}_1} \alpha_k \ket{k} \bigotimes\limits_{j \neq j^*} \ket{{(T_k)}_j}_j
		\end{multlined}
	\right] \ket{1}_{j^*}.
\end{multline}
Hence, if the measurement of the $j^*$th qubit yields $0$, the qubit is properly uncomputed.
If the result is $1$, a phase shift needs to be applied on states corresponding to the indexes $k \in \mathcal{K}_1$.

This uncomputation is successively applied to all the qubits encoding the numbers $T_k$.
Let $t_j$ be the measurement result of the $j$th qubit.
The state after all the measurements is given by
\begin{equation}\label{eq:statekTk}
	 \sum\limits_k \alpha_k \sigma_k \ket{k},
\end{equation}
with $\sigma_k = \prod\limits_{j} {(-1)}^{t_j {(T_k)}_j}$.
We now label
\begin{equation}\label{eq:mathacalK}
\mathcal{K} = \lbrace k \mid \sigma_k = -1 \rbrace.
\end{equation}
In order to recover the desired state, we need to correct selectively the phase of terms $\ket{k}$ for which $k \in \mathcal{K}$.

\begin{figure}[h]
    \mbox{\small \providecommand{\multicontrolgate}[2]{\multimeasure{#1}{#2}}
\providecommand{\controlgate}[1]{\measure{#1}}
\providecommand{\nmultigate}[2]{*+<1em,.9em>{\hphantom{#2}} \POS [0,0]="i",[0,0].[#1,0]="e",!C *{#2},"e"+UR;"e"+UL **\dir{-};"e"+DL **\dir{-};"e"+DR **\dir{-};"e"+UR **\dir{-},"i"}
\providecommand{\ngate}[1]{*+<.6em>{#1} \POS ="i","i"+UR;"i"+UL **\dir{-};"i"+DL **\dir{-};"i"+DR **\dir{-};"i"+UR **\dir{-},"i"}

\Qcircuit @R=0.6em @C=0.3em {
\lstick{\ket{k_{:s}}}&\qw&\qw&\qw&\qw&\qw&{/}\qw&\ustick{s}        \qw&\controlgate{\text{Input }k_{:s}}&   \qw&\qw            &\qw     &\qw                              &\qw     &\controlgate{\text{Input }k_{:s}}&\qw                                           \\
\lstick{\ket{k_{s:}}}&\qw&\qw&\qw&\qw&\qw&{/}\qw&\ustick{w_e+w_m-s}\qw&\qw\qwx                          &   \qw&\qw            &\qw     &\controlgate{\text{Input }k_{s:}}&\qw     &\qw\qwx                          &\qw                                           \\
                     &   &   &   &   &   &      &                     &\ngate{\text{Init unary}}\qwx    &{/}\qw&\ustick{2^s}\qw&\gate{H}&\gate{\oplus F_{k_{s:}}}     \qwx&\gate{H}&\gate{\text{Deinit unary}}\qwx   &                                              \\
} }
    \caption{
    Representation of the four steps proposed in Ref.\,\cite{Gidney2019Windowedquantumarithmetic} to selectively change the phase of components $\ket{k}$ in the state given in~\eqref{eq:statekTk} when the index $k$ belongs to $\mathcal{K}$~\eqref{eq:mathacalK}.
    The central operation is a table lookup with the values $F_{k_{s:}} = \sum\limits_{j=0}^{2^s-1} 2^j \delta(j + 2^s k_{s:})$ where $\delta()$ is the indicator function of $\mathcal{K}$.
    }\label{fig:selective_correct}
\end{figure}
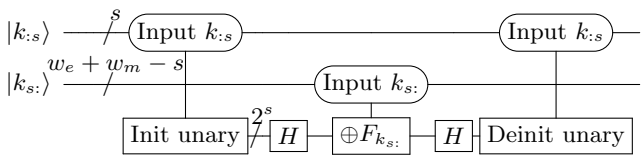

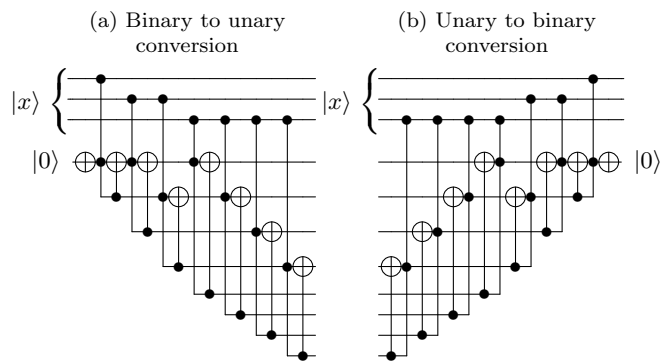
\begin{figure}[h]
\subfloat[Binary to unary conversion]{\label{subfig:unary:init}
	\mbox{\small \Qcircuit @R=0.6em @C=0.1em {
&\qw&\qw             &\qw  &\ctrl{4}&\qw      &\qw     &\qw      &\qw     &\qw      &\qw     &\qw      &\qw     &\qw      &\qw     &\qw      &\qw     &\qw      &\qw                                        \\
&\qw&\qw             &\qw  &\qw     &\qw      &\ctrl{3}&\qw      &\ctrl{4}&\qw      &\qw     &\qw      &\qw     &\qw      &\qw     &\qw      &\qw     &\qw      &\qw                                        \\
&\qw&\qw             &\qw  &\qw     &\qw      &\qw     &\qw      &\qw     &\qw      &\ctrl{2}&\qw      &\ctrl{3}&\qw      &\ctrl{4}&\qw      &\ctrl{5}&\qw      &\qw\inputgroupv{1}{3}{0.7em}{1em}{\ket{x}} \\
&   &                &     &        &         &        &         &        &         &        &         &        &         &        &         &        &         &                                           \\
&   &\lstick{\ket{0}}&\targ&\ctrl{1}&\targ    &\ctrl{2}&\targ    &\qw     &\qw      &\ctrl{4}&\targ    &\qw     &\qw      &\qw     &\qw      &\qw     &\qw      &\qw                                        \\
&   &                &     &        &\ctrl{-1}&\qw     &\qw      &\ctrl{2}&\targ    &\qw     &\qw      &\ctrl{4}&\targ    &\qw     &\qw      &\qw     &\qw      &\qw                                        \\
&   &                &     &        &         &        &\ctrl{-2}&\qw     &\qw      &\qw     &\qw      &\qw     &\qw      &\ctrl{4}&\targ    &\qw     &\qw      &\qw                                        \\
&   &                &     &        &         &        &         &        &\ctrl{-2}&\qw     &\qw      &\qw     &\qw      &\qw     &\qw      &\ctrl{4}&\targ    &\qw                                        \\
&   &                &     &        &         &        &         &        &         &        &\ctrl{-4}&\qw     &\qw      &\qw     &\qw      &\qw     &\qw      &\qw                                        \\
&   &                &     &        &         &        &         &        &         &        &         &        &\ctrl{-4}&\qw     &\qw      &\qw     &\qw      &\qw                                        \\
&   &                &     &        &         &        &         &        &         &        &         &\       &\        &        &\ctrl{-4}&\qw     &\qw      &\qw                                        \\
&   &                &     &        &         &        &         &        &         &        &         &\       &\        &        &         &        &\ctrl{-4}&\qw                                        \\
} }
}
\hfil
\subfloat[Unary to binary conversion]{\label{subfig:unary:deinit}
	\mbox{\small \Qcircuit @R=0.6em @C=0.1em {
&\qw      &\qw     &\qw      &\qw     &\qw      &\qw     &\qw      &\qw     &\qw      &\qw     &\qw      &\qw     &\qw       &\ctrl{4} &\qw  &\qw                &\qw                                        \\
&\qw      &\qw     &\qw      &\qw     &\qw      &\qw     &\qw      &\qw     &\qw      &\ctrl{4}&\qw      &\ctrl{3}&\qw       &\qw      &\qw  &\qw                &\qw                                        \\
&\qw      &\ctrl{5}&\qw      &\ctrl{4}&\qw      &\ctrl{3}&\qw      &\ctrl{2}&\qw      &\qw     &\qw      &\qw     &\qw       &\qw      &\qw  &\qw                &\qw\inputgroupv{1}{3}{0.7em}{1em}{\ket{x}} \\
&         &        &         &        &         &        &         &        &         &        &         &        &          &         &     &                   &                                           \\
&\qw      &\qw     &\qw      &\qw     &\qw      &\qw     &\targ    &\ctrl{4}&\qw      &\qw     &\targ    &\ctrl{2}&\targ     &\ctrl{1} &\targ&\rstick{\ket{0}}\qw&                                           \\
&\qw      &\qw     &\qw      &\qw     &\targ    &\ctrl{4}&\qw      &\qw     &\targ    &\ctrl{2}&\qw      &\qw     &\ctrl{-1} &\qw      &     &                   &                                           \\
&\qw      &\qw     &\targ    &\ctrl{4}&\qw      &\qw     &\qw      &\qw     &\qw      &\qw     &\ctrl{-2}&\qw     &          &         &     &                   &                                           \\
&\targ    &\ctrl{4}&\qw      &\qw     &\qw      &\qw     &\qw      &\qw     &\ctrl{-2}&\qw     &         &        &          &         &     &                   &                                           \\
&\qw      &\qw     &\qw      &\qw     &\qw      &\qw     &\ctrl{-4}&\qw     &         &        &         &        &          &         &     &                   &                                           \\
&\qw      &\qw     &\qw      &\qw     &\ctrl{-4}&\qw     &         &        &         &        &         &        &          &         &     &                   &                                           \\
&\qw      &\qw     &\ctrl{-4}&\qw     &         &        &         &        &         &        &         &        &          &         &     &                   &                                           \\
&\ctrl{-4}&\qw     &         &        &         &        &         &        &         &        &         &        &          &         &     &                   &                                           \\
} }
}
\caption{
    \protect\subref{subfig:unary:init}: representation of the circuit proposed in Ref.\,\cite{Gidney2019Windowedquantumarithmetic} for preparing a copy in a ancillary register of an integer $x$ in a unary representation starting from an encoding of $x$ in a control register in the binary representation.
    The first not operation prepares the first qubit in the ancillary register in state $\ket{1}$.
    The first AND computation writes the result of an AND operation between the first bit of $x$ and the bit $1$ encoded in the first qubit of the ancillary register into the second qubit of the ancillary register.
    In case the state of the latter is $\ket{1}$, the state of the first qubit of the ancillary register is changed to $\ket{0}$.
    The combination of AND and CNOT operations is successively repeated until the desired qubit of the ancillary register is in state $\ket{1}$.
    \protect\subref{subfig:unary:deinit}: representation of the circuit proposed in Ref.\,\cite{Gidney2019Windowedquantumarithmetic} to erase the value in the ancillary register while keeping the integer $x$ into the control register.
    The circuits \protect\subref{subfig:unary:init} and \protect\subref{subfig:unary:deinit} corresponds to the first and third operations needed for the selective phase correction operation presented in \autoref{fig:selective_correct}.
    }\label{fig:unary}
\end{figure}

The selective phase correction is done in four steps~\cite{Gidney2019Windowedquantumarithmetic}, as shown in~\autoref{fig:selective_correct}.
First, the control register which uses $w_e+w_m$~qubits in state $\ket{k}$ is split in two groups.
The first group is made with $s$~qubits in state $\ket{k_{:s}}$.
The second group takes the remaining $w_e+w_m-s$~qubits in state $\ket{k_{s:}}$, such that $\ket{k}=\ket{k_{s:}} \otimes \ket{k_{:s}}$ and $k = k_{:s} + 2^s k_{s:}$.
The second step consists in writing the integer $k_{:s}$ in an ancillary register in the unary representation: a register with $2^s$~qubits representing a number $k_{:s}$ with the state of the qubit number $k_{:s}$ being $\ket{1}$ and all the other qubits in the state $\ket{0}$.
The qubits in state $\ket{k_{s:}}$ and the ancillary qubits are then used as control and target qubits for a lookup circuit where the controlled multi-NOT gates are replaced by controlled multi-$Z$ gates.
Finally, the ancillary register is uncomputed.
The circuit used to initialize the ancillary register in shown in~\autoref{subfig:unary:init}.
The one used to put it back in its initial state is given in~\autoref{subfig:unary:deinit}.

Starting from $x$ encoded in $s$~qubits, the conversion to the unary representation takes $1$~NOT gate, $2^s - 1$~CNOT gates and $2^s - 1$~AND computation.
The conversion back to the binary representation takes $1$~NOT gate, $2^s - 1$~CNOT gates and $2^s - 1$~AND uncomputation.
Given that $k$ is encoded in $w_e+w_m$~bits, and that a choice $s = \floor{\frac{w_e + w_m}{2}}$ is judicious to minimize the number of gates, the change of phase of components $\ket{k}$ takes
$2^{\floor{\frac{w_e + w_m}{2}}+1} + 4$~1-qubit gates, $2^{w_e + w_m - 1} + 2^{\floor{\frac{w_e + w_m}{2}}+1} + 2^{\ceil{\frac{w_e + w_m}{2}}} - 4$~CNOTs and $2^{\floor{\frac{w_e + w_m}{2}}} + 2^{\ceil{\frac{w_e + w_m}{2}}} - 3$~ANDs
($1$~NOT gate, $2^{\floor{\frac{w_e + w_m}{2}}} - 1$~CNOT gates and $2^{\floor{\frac{w_e + w_m}{2}}} - 1$~AND computation for the unary conversion,
$2 \times 2^{\floor{\frac{w_e + w_m}{2}}}$~Hadamard gates around the table lookup,
$2$~NOT gates, $2^{\ceil{\frac{w_e + w_m}{2}}} - 2 + 2^{w_e + w_m-1}$~CNOT gates, $2^{\ceil{\frac{w_e + w_m}{2}}} - 2$~AND computations and uncomputations for the lookup circuit and $1$~NOT gate, $2^{\floor{\frac{w_e + w_m}{2}}} - 1$~CNOT gates and $2^{\floor{\frac{w_e + w_m}{2}}} - 1$~AND uncomputation for the binary conversion).
Including the additional $n$~Hadamard gates and $n$~measurements on $T_k$, we conclude that the table unlookup takes $2^{\floor{\frac{w_e + w_m}{2}}+1} + n + 4$~1-qubit gates, $2^{w_e + w_m - 1} + 2^{\floor{\frac{w_e + w_m}{2}}+1} + 2^{\ceil{\frac{w_e + w_m}{2}}} - 4$~CNOTs and $2^{\floor{\frac{w_e + w_m}{2}}} + 2^{\ceil{\frac{w_e + w_m}{2}}} - 3$~ANDs.

\subsection{Standard adder}
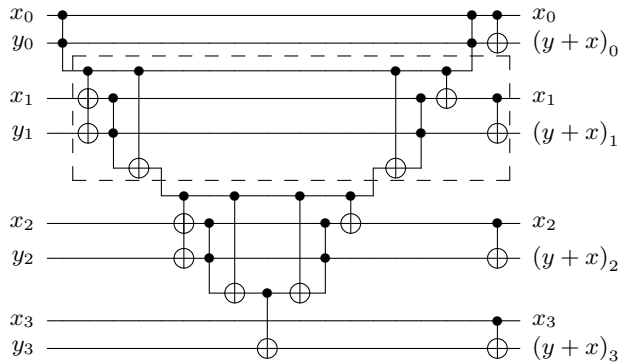
\begin{figure}[h]
	\hspace*{-2.5em}\mbox{\small \Qcircuit @R=0.6em @C=0.5em {
	    &\lstick{x_0}&\ctrl{1} &\qw     &\qw     &\qw     &\qw &\qw     &\qw     &\qw     &\qw     &\qw     &\qw     &\qw     &\qw    &\qw     &\qw     &\qw     &\ctrl{1}&\ctrl{1}&\rstick{x_0}\qw       \\
	    &\lstick{y_0}&\ctrl{1} &\qw     &\qw     &\qw     &\qw &\qw     &\qw     &\qw     &\qw     &\qw     &\qw     &\qw     &\qw    &\qw     &\qw     &\qw     &\ctrl{1}&\targ   &\rstick{{(y+x)}_0}\qw \\
	    &            &         &\ctrl{2}&\qw     &\ctrl{3}&\qw &\qw     &\qw     &\qw     &\qw     &\qw     &\qw     &\qw     &\qw    &\ctrl{3}&\qw     &\ctrl{1}&\qw     &        &                      \\
	    &\lstick{x_1}&\qw      &\targ   &\ctrl{1}&\qw     &\qw &\qw     &\qw     &\qw     &\qw     &\qw     &\qw     &\qw     &\qw    &\qw     &\ctrl{1}&\targ   &\qw     &\ctrl{1}&\rstick{x_1}\qw       \\
	    &\lstick{y_1}&\qw      &\targ   &\ctrl{1}&\qw     &\qw &\qw     &\qw     &\qw     &\qw     &\qw     &\qw     &\qw     &\qw    &\qw     &\ctrl{1}&\qw     &\qw     &\targ   &\rstick{{(y+x)}_1}\qw \\
	    &            &         &        &        &\targ   &\qw &        &        &        &        &        &        &        &       &\targ   &\qw     &        &        &        &                      \\
	    &            &         &        &        &        &\qwx&\ctrl{2}&\qw     &\ctrl{3}&\qw     &\ctrl{3}&\qw     &\ctrl{1}&\qw\qwx&        &        &        &        &        &                      \\
	    &\lstick{x_2}&\qw      &\qw     &\qw     &\qw     &\qw &\targ   &\ctrl{1}&\qw     &\qw     &\qw     &\ctrl{1}&\targ   &\qw    &\qw     &\qw     &\qw     &\qw     &\ctrl{1}&\rstick{x_2}\qw       \\
	    &\lstick{y_2}&\qw      &\qw     &\qw     &\qw     &\qw &\targ   &\ctrl{1}&\qw     &\qw     &\qw     &\ctrl{1}&\qw     &\qw    &\qw     &\qw     &\qw     &\qw     &\targ   &\rstick{{(y+x)}_2}\qw \\
	    &            &         &        &        &        &    &        &        &\targ   &\ctrl{2}&\targ   &\qw     &        &       &        &        &        &        &        &                      \\
	    &\lstick{x_3}&\qw      &\qw     &\qw     &\qw     &\qw &\qw     &\qw     &\qw     &\qw     &\qw     &\qw     &\qw     &\qw    &\qw     &\qw     &\qw     &\qw     &\ctrl{1}&\rstick{x_3}\qw       \\
	    &\lstick{y_3}&\qw      &\qw     &\qw     &\qw     &\qw &\qw     &\qw     &\qw     &\targ   &\qw     &\qw     &\qw     &\qw    &\qw     &\qw     &\qw     &\qw     &\targ   &\rstick{{(y+x)}_3}\qw
	    \gategroup{3}{4}{6}{20}{1em}{--}
}
 }
\caption{
	Adder modulo $2^4$ from~\cite{GidneyQ2018Halvingcostquantum}, using the same notations as in \autoref{fig:comparison}.
	The building block (boxed) is repeated two times, for the qubits numbers $1$ and $2$, while the first and last use a simplified subcircuit.
	}\label{fig:adder}
\end{figure}

As we use the coset representation of integers with windowed arithmetic operations, a circuit for unconditional addition modulo a power of two is sufficient to implement a modular addition.
The adder we use, which is described in~\cite{GidneyQ2018Halvingcostquantum} and optimized from~\cite{Moulton2004newquantumripple} for use with $T$ gates, is presented in \autoref{fig:adder}.
It is thrifty in gate number and ancillary qubits, at the cost of being deeper than other circuits~\cite{Draper2000AdditionQuantumComputer,SvoreQIaC2006logarithmicdepthquantum}, which is not a disadvantage for our architecture.

As presented in \autoref{subfig:exponentiation:lookup_add}, the adder needs to add a number $T_k$ taking $n$~qubits into a register with $n+m$~qubits.
To achieve this, either the first register for $T_k$ is extended with qubits in the $\ket{0}$ state, either we use carry propagation blocs for the last qubits.
Such blocs are identical to the ones of semi-classical adder with classical input $0$; see~\cite[Fig.\,17]{Babbush2020CompilationFaultTolerant} for an example of such a circuit.
For gate counting, the first solution is taken into account.

The cost of the addition circuit (\autoref{fig:adder}) is $6 (n + m) - 9$~CNOT gates and $n + m -1$~AND computations and uncomputations.  

\subsection{Cost estimation}
In summary, the parameters of the logical circuit for computing the modular exponentiation are
\begin{enumerate}
\item[$n$] number of bits of the exponentiated number $g$
\item[$n_e$] number of bits of the exponent $e$
\item[$w_e$] window size for the exponentiation
\item[$w_m$] window size for the multiplication
\item[$m$] number of qubits added by the coset representation
\end{enumerate}
The aim of this subsection is to give an estimate of the number of gates needed to implement this circuit.
In order to keep the evaluation independent of the error correction choice, we express the cost in terms of the number of 1-qubit, 2-qubit gates and AND computation and uncomputation~\footnote{Due to the measurement-based uncomputation, it is often more efficient to implement the non-Clifford operations through AND computation.
In case of direct implementation of Toffoli gates, the circuit cost could be slightly reduced.}.

The modular exponentiation consists in $n_e/w_e$~multiplications, each multiplication using $2$~product addition and a swap and each product addition is implemented with $(n+m)/w_m$~lookups, additions and unlookups.
Note that the swap operation is realized by simply relabeling the register, hence is for free.
According to the counts obtained from previous subsections, the cost of the exponentiation is dominated --- in the limit $n \to \infty$, $n_e = \bigO{n}$, $w_e$ and $w_m$ constant --- by: $2 \frac{n_e (n+m) n}{w_e w_m}$~1-qubit gates, $\left(2^{w_e + w_m}n + 12(n+m)\right) \frac{n_e (n+m)}{w_e w_m}$~CNOTs, and $2 \frac{n_e {(n+m)}^2}{w_e w_m}$~AND computations and uncomputations (translatable into $4 \frac{n_e {(n+m)}^2}{w_e w_m}$~Toffoli gates).
Note that when considering the universal gate set $T$, $S$, $H$, $X$, $Y$, $Z$, CNOT, controlled-Z and their conjugate, according to Fig.\,4 of Ref.\,\cite{NevenPRX2018EncodingElectronicSpectra} the AND computation and uncomputation costs in average $8$~1-qubit gates and $3.5$~2-qubit gates.
The total cost of the exponentiation is hence given at the leading order by $2 \frac{n_e (n+m) n}{w_e w_m} \left(9n + 8m\right)$~1-qubit gates and $\left(2^{w_e + w_m}n + 19(n+m)\right) \frac{n_e (n+m)}{w_e w_m}$~2-qubit gates.
In the code used to compute the required resources and find the optimal parameters, the complete formula have been used~\cite{Note1}.

\section{Error correction}\label{appendix:err_correction}
This appendix is dedicated to 3D gauge color codes.
The first subsection is dedicated to the principle of subsystem codes.
The second subsection describes the geometrical structure of 3D gauge color codes.
The last subsection provides a detailed description of the cut of the code structure that is used to process and correct the logical qubits.

\subsection{Subsystem codes}
Subsystem stabilizer codes~\cite{PoulinPRL2005StabilizerFormalismOperator} are defined by three subgroups of the Pauli group: the stabilizer, gauge and logical (also designated as \emph{bare logical} in~\cite{BombinNJoP2015Gaugecolorcodes}) operator groups, such that the stabilizer group is the center of the gauge group up to phases, $i\identite$ is included in the gauge group, the operators from the gauge and logical groups commutes, and the normalizer of the stabilizer group is the product of gauge and logical groups.
We invite the reader to look at Refs.\,\cite{PoulinPRL2005StabilizerFormalismOperator,BombinNJoP2015Gaugecolorcodes} for an explicit construction of those groups from canonical generators of the Pauli group.
The stabilizer group plays the standard role of stabilizers, \latin{i.e.\@} divides the total Hilbert space $\mathcal{H}$ into a direct sum of orthogonal subspaces $C \oplus C^\perp$ where $C$ --- the stabilized subspace --- corresponds to the eigenspace $+1$ of all stabilizers.
The gauge and logical groups decompose the stabilized subspace $C$ into a tensor product of the logical qubits space $A$ and the gauge qubits space $B$~\cite{LloydPRL2004QuantumTensorProduct}, that is, the Hilbert space is decomposed as
\begin{equation*}
\mathcal{H} = \underbrace{(A \otimes B)}_{C} \oplus C^\perp.
\end{equation*}
The gauge group acts trivially on the logical qubits and is the Pauli group of the gauge qubits while the logical group acts trivially on the gauge qubits and is the Pauli group of the logical qubits (up to phases).
This ensures that gauge operator measurements don't modify the logical qubits.

A gauge fixing operation consists in switching from a code to another one such that the new stabilizer group includes the original one while being included into the original gauge group, while keeping unmodified the logical group.
The decomposition associated to the original code
\begin{equation*}
\mathcal{H} = \underbrace{(A \otimes B)}_{C} \oplus C^\perp
\end{equation*}
then becomes of the form
\begin{equation*}
\mathcal{H} = (A \otimes B') \oplus \underbrace{(A \otimes B'') \oplus C^\perp}_{C'^\perp}
\end{equation*}
where $B'$ is the new gauge qubit space.
As a consequence, a valid code-word for this new code is also valid for the initial one.
The passage of the latter to the new code is done by measuring the generators of the gauge group, the results of these measurements giving the correction to apply on $B' \oplus B''$ to remove the components on $B''$.

For 3D gauge color codes, code switching allows a transversal error-corrected implementation of a universal set of gates~\cite{BombinNJoP2015Gaugecolorcodes}.

\subsection{Code geometrical structure}
The geometrical structure of the 3D gauge color codes is described in detail in Section\,3.1 of~\cite{BombinNJoP2015Gaugecolorcodes}.
It takes a large tetrahedron, itself decomposed into elementary tetrahedrons, see \autoref{fig:slice} for an example.
Four extra points ($v_i, i\in \{1,2,3,4\}$) are then added outside the large tetrahedron, one point in front of each facet of the large tetrahedron.
Elementary tetrahedrons are finally added between those extra points and the vertices at the surface of the large tetrahedron, see Fig.\,4b of~\cite{Svore2021costuniversalitycomparative} for an illustration.
The vertices of elementary tetrahedrons are colored with 4 different colors such that adjacent vertices get a different color.
Each elementary tetrahedron represents a physical qubit.

The measured operators are the gauge generators for the code used to implement the $H$ and CNOT gates --- the $(1,1)$ code (see~\cite{BombinNJoP2015Gaugecolorcodes}).
These generators are described by the edges: each operator is the product of $X$ or $Z$ operators of the elementary tetrahedrons adjacent to a given edge (each operator implies up to 6 physical qubits).

The stabilizer generators of the $(1,1)$ and $(1,2)$ codes (the $(1,2)$ code refers to the code used to implement the $T$ gate~\cite{BombinNJoP2015Gaugecolorcodes}) which are described by the vertices and edges, are deduced from the values of measured operators.
More precisely, the operator corresponding to a vertex can be written as the product of the operators corresponding to edges starting at the given vertex and ending on vertices of a common color.
Three choices of color are possible, allowing one to recover in three different ways an operator corresponding to a vertex.
This redundancy can be used for achieving fault-tolerant error correction in only one measurement of the (gauge) operators related to the edges~\cite{BombinNJoP2015Gaugecolorcodes}.

Let $n_\text{code}$ be the index of the code which is the number of vertices of the same color on one edge of the large tetrahedron (denoted as $n$ in~\cite{BombinNJoP2015Gaugecolorcodes}).
The code distance is given by $d = 2 n_{\text{code}} + 1$, and the number of physical qubits is $ 1 + 4n_{\text{code}} + 6n_{\text{code}}^2 + 4 n_{\text{code}}^3 = \frac{d^3 + d}{2}$~\cite{BombinNJoP2015Gaugecolorcodes,Svore2021costuniversalitycomparative}.

\subsection{Slicing of the code structure}
\begin{figure}[h]
    \includegraphics[width=0.5\linewidth]{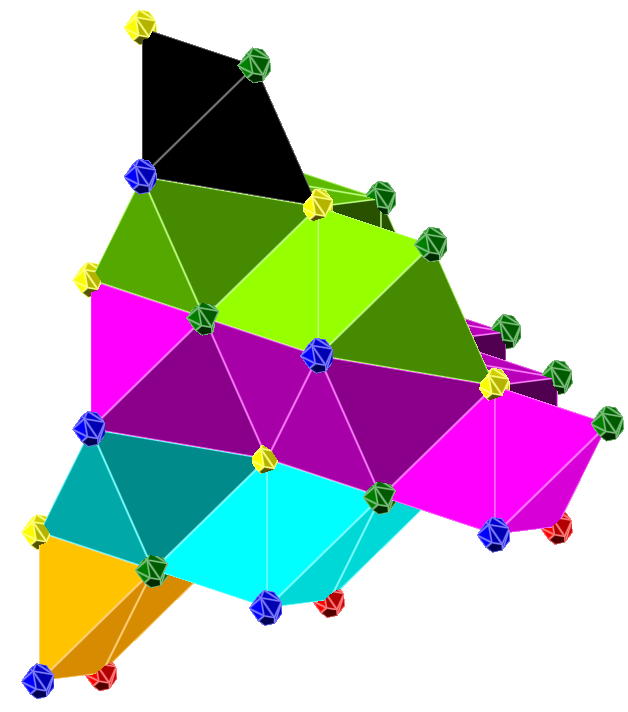}
    \caption{
        Code geometrical structure for $n_\text{code} = 3$ (without the extra points ($v_i, i\in \{1,2,3,4\}$).
        Each slice has been represented with a specific color.
        The larger slice is with the magenta elementary tetrahedrons.
        The figure shows that the maximum number of slices involved in an operator corresponding to an edge is $2$.
    }\label{fig:slice}
\end{figure}

To process the information, the code structure is decomposed into slices, each slice being map successively into the 2D processor.
While several cuts in slices are possible, we choose slices orthogonal to two faces (see \autoref{fig:slice}).
The processor need to be sized to fit in the larger slice, that join the edge not included into any of the two faces to the middle of the opposing edge --- the magenta slice in \autoref{fig:slice}.

With the lattice described in Ref.\,\cite{BombinNJoP2015Gaugecolorcodes}, the central slice corresponds to the elementary tetrahedrons
for which all vertices coordinates satisfy $x+z = n_{\text{code}}-2$ or $x+z = n_{\text{code}}-1$
(the elementary tetrahedrons between the two plans defined by the previous equations).
Note that the number of slices is given by $d-2$.

\begin{figure}[h]
\subfloat[First set]{\label{subfig:tranche:petit}
	\includegraphics{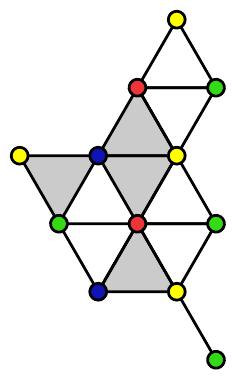}
}
\hfil
\subfloat[Second set]{\label{subfig:tranche:grand}
	\includegraphics{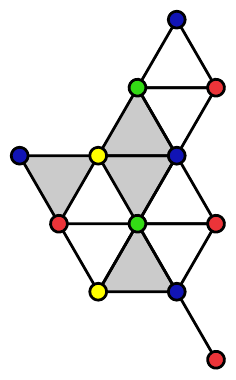}
}
\hfil
\subfloat[Third set]{\label{subfig:tranche:milieu}
	\includegraphics{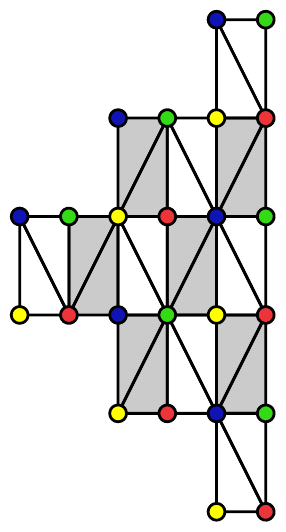}
}
    \caption{
        Decomposition of the central slice for $n_\text{code} = 3$ (magenta slice in the tetrahedron presented in \autoref{fig:slice}).
        Each subfigure corresponds to a set of elementary tetrahedrons of the central slice, seen from different point of views.
        On \protect\subref{subfig:tranche:petit} and \protect\subref{subfig:tranche:grand}, each triangle corresponds to an elementary tetrahedron.
        On \protect\subref{subfig:tranche:milieu} each small rectangle correspond to an elementary tetrahedron.
    }\label{fig:tranche}
\end{figure}

The number of elementary tetrahedrons included in this slice is counted by considering three tetrahedron sets, see \autoref{fig:slice} and \autoref{fig:tranche}.
Two sets correspond to the elementary tetrahedrons having a facet at the interplay between two slices (\autoref{subfig:tranche:petit} (\autoref{subfig:tranche:grand}) is associated to the elementary tetrahedrons with one facet are the interplay between the magenta slice and the green (cyan) slice).
The last set is associated to the elementary tetrahedrons having no facet at the interplay between two slices (\autoref{subfig:tranche:milieu}).
One can check that the two first sets include
$\sum\limits_{k=1}^{2n_{\text{code}-2}} k = 2n_{\text{code}}^2 - 3 n_{\text{code}} + 1$~elementary tetrahedrons while the last set has $(2n_{\text{code}}-1) + \sum\limits_{k=0}^{n_{\text{code}}-2} 2 (2k+1) = 2n_{\text{code}}^2 - 2 n_{\text{code}} + 1$~elementary tetrahedrons.
They are $16 n_{\text{code}} - 2$~additional elementary tetrahedrons resulting from the $4$~added points in the construction of the code.
In total, the maximum number of elementary tetrahedron for one slice of the code structure is $6n_{\text{code}}^2 + 8n_{\text{code}} + 1$.
Since we consider a processor that can process up to two slices (associated to two different logical qubits) and accounting for the ancillary subsystems needed to measure the gauge generators by a simple factor of two, we obtain the number of physical qubits in the processor specified in the main text.
For more details, see the ancillary file \verb|tetrahedron_3_bis.scad|~\cite{Note1}, where each tetrahedron color corresponds to a given slice, the larger being the magenta one.

\subsection{Threshold of 3D gauge color codes}\label{appendix:code:threshold}

The value of the threshold for 3D gauge color codes has been evaluated in a few references that we now discuss.

In order to clarify on the context, let us first remind that there are three main definitions of error-rate threshold used for stabilizer codes in the literature: code-capacity, phenomenological and circuit-level.
Code-capacity thresholds assume perfect measurements of stabilizers.
Phenomenological thresholds model faulty-measurements as bit-flip errors on stabilizer measurement outcomes.
Circuit-level thresholds model errors occurring at any stage of stabilizer measurement circuits.

In Ref.~\cite{BrowneNC2016Faulttoleranterror} a clustering decoding scheme is presented and
by including a phenomenological noise to the measurement outputs, the authors estimate a code-capacity threshold of \SI{0.46}{\percent} and phenomenological threshold to about \SI{0.31}{\percent}, suggesting an upper bound on the circuit-level threshold.
Note however that the underlying lattice considered in Ref.~\cite{BrowneNC2016Faulttoleranterror} (cubic lattice) is different from the one we have considered (body centered cubic lattice (bcc)).

A more recent decoding algorithm is presented in~\cite{Kubica2018ABCsColorCode,Delfosse2019Efficientcolorcode} using the bcc lattice, but the authors give an estimate of the code capacity threshold of \SI{0.77}{\percent} only.
By the way, a slightly better code capacity threshold of \SI{0.80}{\percent} has been estimated in Ref.~\cite{Svore2021costuniversalitycomparative} under the same assumptions.

Finally, statistical arguments have be used in Ref.~\cite{SvorePRL2018ThreeDimensionalColor} to estimate code-capacity threshold of 3D gauge color codes with ideal decoding to around \SI{1.9}{\percent}.
This suggests that an appropriate decoder could significantly improve the value of the code-capacity threshold and hence of the phenomenological and circuit-level thresholds.

\begin{figure}[h]
\centering
\includegraphics[width=0.5\textwidth]{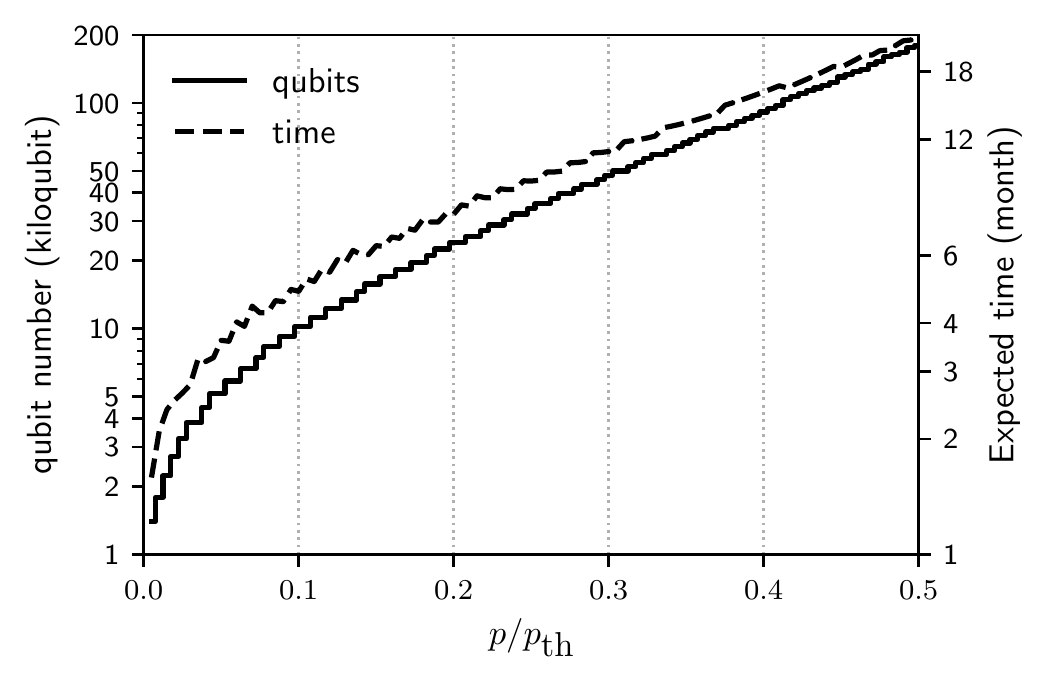}
\caption{Number of qubits in the processor and run-time to factor of \num{2048}-bit RSA integers in function of the ration between the physical qubit error and the fault-tolerant code threshold.}\label{fig:ressources_noise_ratio}
\end{figure}

Since we believe that the determination of the circuit-level threshold goes beyond the scope of this work, the run-time and resource needed to factor a \num{2048}-bit RSA integer are given in the main text under the assumption of a threshold of \SI{0.75}{\percent}.
Since this choice is somehow arbitrary, we give the evolution of run-time and resource as a function of the threshold in \autoref{fig:ressources_noise_ratio}.
More precisely, they are given as a function of the ratio $p/p_{\text{th}}$ between the physical error probability per cycle $p$ and the fault-tolerant threshold $p_{\text{th}}$ which is the only relevant quantity at first order.
For $p_{\text{th}}=\SI{0.75}{\percent}$ and an error probability per cycle and per physical qubit of $10^{-3}$, this ratio $p/p_{\text{th}}$ is given by $\approx 0.13$.

We emphasize that the value of the threshold for 3D gauge color codes does not change the take home message of the whole paper, namely that the use of a quantum memory in quantum computing strongly reduces the number of qubits in the processing unit.
Even when considering for example a circuit-level threshold of \SI{0.2}{\percent} and a error probability per operation of $10^{-3}$, the use of a quantum memory reduces the number of qubits by two orders of magnitude in the processor compared to an architecture without memory for factoring \num{2048}-bit RSA integers (the same conclusion holds when considering the standard approach using surface code, see \autoref{appendix:subsec:decouplage}).

\section{Results and possible improvements}\label{appendix:results}
We presented in the main text the resources needed to factor \num{2048}-bit RSA integers corresponding to the most common RSA key size.
In the first subsection of this appendix, we discuss the factorization of RSA integers of various sizes.
The second subsection is dedicated to a discussion on ways to reduce the run-time to factor RSA integers and in particular, on the trade-off between the number of physical qubits in the processor and the run-time.

\subsection{Optimal parameters to factor $n$-bit RSA integers}
The resources and parameters needed to factor RSA integers encoded in $n$~bits are specified in \autoref{table:parametres}.
In particular, we consider the factorization of RSA integers with $n=6$~bits, the number of bits needed to factor $35$.
We also consider $n=829$ which corresponds to the largest RSA integer factorized so far~\cite{Zimmermann2020FactorizationRSA250}.
\begin{table*}
\begin{tabular}{S[table-figures-integer=4]S[table-figures-integer=4]|S[table-figures-integer=2]S[table-figures-integer=1]S[table-figures-integer=1]S[table-figures-integer=2]|S[table-figures-integer=5]c|S[table-figures-integer=4]S[table-figures-integer=9]S[table-figures-integer=8]S[table-figures-integer=2]c}
	{$n$}                         &{$n_e$}                       &{$m$}                         &{$w_e$}                       &{$w_m$}                       &{$d$}                         &{$n_{\text{qubits}}$}         &{$t_{\text{exp}}$}            &{logical qubits}              &{total modes}                 &{spatial modes}               &{temporal modes}              &{all memory correction}       \\ \hline
	6                             &6                             &4                             &3                             &2                             &7                             &316                           &\SI{1}{\minute}               &38                            &6650                          &3002                          &5                             &\SI{95}{\micro\second}        \\
	8                             &9                             &8                             &3                             &2                             &13                            &1060                          &\SI{2}{\second}               &58                            &64090                         &15370                         &11                            &\SI{319}{\micro\second}       \\
	16                            &21                            &11                            &3                             &2                             &17                            &1796                          &\SI{10}{\second}              &99                            &244035                        &44451                         &15                            &\SI{742}{\micro\second}       \\
	128                           &189                           &19                            &3                             &3                             &29                            &5156                          &\SI{50}{\minute}              &571                           &6971339                       &736019                        &27                            &\SI{8}{\milli\second}         \\
	256                           &381                           &21                            &3                             &3                             &33                            &6660                          &\SI{7}{hours}                 &1089                          &19585665                      &1813185                       &31                            &\SI{17}{\milli\second}        \\
	512                           &765                           &24                            &3                             &3                             &37                            &8356                          &\SI{2}{days}                  &2122                          &53782090                      &4432858                       &35                            &\SI{37}{\milli\second}        \\
	829                           &1242                          &26                            &3                             &3                             &41                            &10244                         &\SI{11}{days}                 &3396                          &117097476                     &8697156                       &39                            &\SI{66}{\milli\second}        \\
	2048                          &3029                          &30                            &3                             &3                             &47                            &13436                         &\SI{177}{days}                &8284                          &430229540                     &27825956                      &45                            &\SI{186}{\milli\second}       \\
\end{tabular}
\caption{For different integer sizes $n$ and corresponding exponent size $n_e$ ($\sim 1.5 n$), the table presents the optimal set of parameters, processor size and computation run-time, and the memory requirements.}\label{table:parametres}
\end{table*}

\subsection{Trade-off between qubits and run-time}
We have estimated that an average run-time of \SI{177}{days} is needed to factor a \num{2048}-bit RSA number.
There are several ways to reduce this number, most of them coming at the cost of using more qubits in the processor.
The items below present several ways separately.
\begin{itemize}
\item Due to the tetrahedral geometry of the code structure, only one third of the processor qubits are used during the error-correction steps in average.
    A factor 3 in time could thus be saved by making use of them.
\item The logical circuit can be parallelized in several ways, giving a speed-up roughly proportional to the increase in qubit numbers in the processor.
	More precisely:
	\begin{itemize}
	\item Some operations in the adder can be parallelized (see \autoref{fig:adder}).
	    The controlled NOT operations aligned vertically can be applied at the same time.
	\item The run-time is dominated by the time spent to implement the CNOT gates of the quantum lookup circuit (see \autoref{fig:qrom}) and they are easily parallelizable.
		A full parallelization, would reduce the factorization of \num{2048}-bit RSA integers to about \num{27}~days, at the cost of using about \num{12}~million qubits in the processor.
	\item Oblivious carry runways allows parallelization of the adders~\cite{Gidney2019Approximateencodedpermutations}.
	\item Other type of adders could exploit further parallelizations, for instance lookahead adders~\cite{SvoreQIaC2006logarithmicdepthquantum}.
	\item During a product-addition operation, the different additions can be parallelized by computing separately partial sums.
	\item During the exponentiation, the different multiplications can be parallelized by computing separately partial products.
	\end{itemize}
\item The qubit number can be reduced using another slicing of the code structure, at the cost of a longer computation time.
    For example, if one chooses to cut the tetrahedron by slices parallel to a facet of this tetrahedron, we estimate that a \num{2048}-bit RSA integer could be factorized with \num{6628}~qubits in the processor and \num{354}~days.
\end{itemize}

\subsection{Decoupling the gain from 3D gauge color code and multimode memory}\label{appendix:subsec:decouplage}
Two new design elements have been proposed in this manuscript, the use of 3D Gauge color codes and an architecture using a multi-mode memory.
We here separate them out and get insight into the improvements from each.

The main motivation to use 3D gauge color codes is to get rid of the magical state factory needed for implementing non-Clifford gates in surface code.
However, the transversality of T gate on 3D gauge color codes is strongly linked with the dimensionality, and 2D color codes can't directly achieve it~\cite{KoenigPRL2013ClassificationTopologicallyProtected,BombinNJoP2015Gaugecolorcodes}.
There is no direct way to make use of a 3D color code on a 2D grid.

The main advantage brought by the memory is to unload qubits from the processing unit to the memory.
Using a memory in the standard approach for example (2D grid and surface code), we estimate that a RSA-\num{2048} integer can be factorized with a 2D surface code in about 68~days using a memory that can store up to 5~million modes and a processor with \num{184}~thousand qubits, \num{180}~thousand being dedicated to the magical state factory and \num{4}~thousand to the logical qubits on the processor.
The additional reduction in the processor size in our approach comes from the fact that there is no need for magic state distillation in the use 3D gauge color codes.
The number of qubits in the processor is kept small because the qubits are released from the memory and process slice by slice.

\section{Memory requirements to factor RSA-2\,048 integers}\label{appendix:memoire}
We would like to first emphasize that the main objective of our project was to evaluate accurately the performance of an architecture in which unprocessed qubits are stored in a quantum memory.
The standard approach suffers from the need of millions of individually controlled qubits and several research entities are dedicating large teams of engineers to tackle this challenge.
We have shown through Shor's algorithm that the use of a quantum memory reduces significantly the number of qubits in the processor though a significant change in the way the information is processed and protected against errors.
Our results hence provide a solution to an engineering problem and turns it into a physics problem: the implementation of a faithful and multimode memory.
Before discussing the requirements on the memory in detail, let us clearly define the notion of multimode memory~\cite{GisinPRL2007QuantumRepeatersPhoton}.

From an algorithm point of view, ``spatial modes'' are stored modes that can be accessed in constant time, while ``temporal modes'' can only be sequentially recoverable (first stored, first release).
In the proposed implementation based on spin-echo, temporal modes correspond to different time slots, photons arriving in different time bins being remitted sequentially after spin refocusing.
Spatial modes correspond to either different spatial (transverse) modes of a cavity or to different cavities (with possibility to combine both).
As discussed in the main text, it is possible to use temporal multiplexing only, at the cost of increasing the run-time.

We now estimate that the factorization of RSA-\num{2048} integers with the proposed architecture would take a memory with the following characteristics:
\begin{itemize}
\item A large multimode capacity to store 28~million spatial modes, each spatial mode being used to store 45~temporal modes.
	We stress that the number of modes in the memory has not been optimized (only the number of qubits in the processor and the run-time are optimized).
	Note also that different choices of processing and error-correction protocols may lead to compromises between the numbers of processing qubits and multimode capacity, if needed.
	For example, we estimate that RSA-\num{2048} integers can be factorized in \SI{68}{days} with a 2D surface code using a memory that can store up to 5~million modes and a processor with \num{184}~thousand qubits in the processor.
\item Storage time greater than \SI{186}{\milli\second}.
	More precisely, we estimate that the maximum storage time between two readouts of the same qubits is less than 2~hours.
	A memory with a storage time of at least two hours is however not necessary as error-correction steps can be implemented periodically at the cost of increasing the run-time.
	Error correction of all the qubits stored in the memory is estimated to take \SI{186}{\milli\second} with a processor having \num{13436}~qubits, meaning that the storage time simply needs to be longer than \SI{186}{\milli\second}.
	Applying a correction every second for example would increase the run-time by about \SI{23}{\percent}.
\item Error probability for a transfer to memory, storage, and retrieval less than \SI{0.1}{\percent}.
	Note that this requirement for a complete cycle of write/read from memory is likely very conservative.
	Indeed, the threshold value of error correction is mainly determined by the errors happening during the stabilizer measurements.
	We thus conjecture that the error correction could handle higher error rate for those specific operations.
	The effect of this strongly dissymmetric noise between the memory/processor operations is still under investigation, and we choose to stick to the conservative hypothesis for this article.
\item The information stored in a given memory mode can be mapped to 3~qubits of the processor: two for the 2-qubit gates (depending on whether the physical qubit is the logical control or target qubits) and one for the error correction and 1-qubit gates.
	No need for an all to all connectivity.
\end{itemize}

\section{Realization combining a rare-earth doped solid and a superconducting resonator}\label{appendix:realization}
For implementing a multimode memory with a spin-echo technique, materials doped with rare-earth, such as Erbium \ch{Er^{3+}} provide an appealing example since these ions have doubly-degenerate Zeeman states which split when an external magnetic field is applied.
Several manuscripts have reported on the successful coupling between the crystal \ch{Er^{3+}: Y2 Si O5} and a superconducting microwave resonator~\cite{UstinovPRB2011Ultralowpowerspectroscopy, WilsonJoPBAMaOP2012Couplingerbiumspin, BushevPRL2013AnisotropicRareEarth}.
Ref.\,\cite{BushevPRL2013AnisotropicRareEarth} in particular reported on the strong coupling with a collective coupling rate $g\sqrt{\bar{N}}=\SI[parse-numbers=false]{2\pi \times 34}{\mega\hertz}$ and an inhomogeneous linewidth $\Gamma=\SI[parse-numbers=false]{2 \pi \times 12}{\mega\hertz}$.
This results in a very high absorption coefficient  $\alpha= \SI{4.0}{\per\meter}$.
If we assume a $L = \lambda/2$ cavity, unit absorption and re-emission efficiencies are obtained if the quality factor is $Q= F = 2\pi/(\alpha \lambda) \approx 26$  for a \SI{5}{\giga\hertz} cavity.
In this low-Q regime, $\kappa \gg \Gamma$ and a coherence time of a few hundreds of microseconds would translate into a multimode capacity of a few tens of modes.
By working with crystals having lower doping concentrations, the coherence time can be significantly increased~\cite{Liu2021Twentymillisecondelectron}, while still reaching the impedance matching point with low-Q resonators.
In this case, a few thousand modes might realistically be stored very efficiently.
Rare-earth doped materials is not the only option and other candidates such as negatively charged nitrogen vacancy color centers in diamond~\cite{BertetPRL2011HybridQuantumCircuit} or bismuth donors in silicon~\cite{BertetPRL2020MultimodeStorageQuantum} may be even more promising.

\bibliography{factorisation_memoire}

\end{document}